\documentclass[twocolumn,english,aps,floatfix]{revtex4}
\usepackage[T1]{fontenc}
\usepackage[latin9]{inputenc}
\usepackage{amsbsy}
\usepackage{graphicx}

\makeatletter

\usepackage{graphics}

\@ifundefined{definecolor}
 {\usepackage{color}}{}

\usepackage{bm}

\usepackage{babel}

\usepackage{babel}

\usepackage{babel}

\usepackage{babel}

\makeatletter
\makeatletter
\makeatletter
\makeatletter
\@ifundefined{definecolor}
 {\@ifundefined{definecolor}
 {\@ifundefined{definecolor}
 {\@ifundefined{definecolor}
 {
}{}
}{}
}{}
}{}
\makeatother
\makeatother
\makeatother
\makeatother

\makeatother

\usepackage{babel}

\begin{document}

\title{Manifestations of nematic degrees of freedom in the magnetic, elastic,
and superconducting properties of the iron pnictides}

\author{Rafael M. Fernandes}

\affiliation{Department of Physics, Columbia University, New York, New York 10027,
USA }

\affiliation{Theoretical Division, Los Alamos National Laboratory, Los Alamos,
NM, 87545, USA}

\author{Jörg Schmalian}

\affiliation{Institute for Theory of Condensed Matter Physics and Center for Functional
Nanostructutes, Karlsruhe Institute of Technology, Karlsruhe, 76131,
Germany }

\date{\today }

\begin{abstract}
We investigate how emergent nematic order and nematic fluctuations
affect several macroscopic properties of both the normal and superconducting
states of the iron pnictides. Due to its magnetic origin, long-range
nematic order enhances magnetic fluctuations, leaving distinctive
signatures in the spin-lattice relaxation rate, the spin-spin correlation
function, and the uniform magnetic susceptibility. This enhancement
of magnetic excitations is also manifested in the electronic spectral
function, where a pseudogap can open at the hot spots of the Fermi
surface. In the nematic phase, electrons are scattered by magnetic
fluctuations that are anisotropic in momentum space, giving rise to
a non-zero resistivity anisotropy whose sign changes between electron-doped
and hole-doped compounds. We also show that due to the magneto-elastic
coupling, nematic fluctuations soften the shear modulus in the normal
state, but harden it in the superconducting state. The latter effect
is an indirect consequence of the competition between magnetism and
superconductivity, and also causes a suppression of the orthorhombic
distortion below $T_{c}$. We also demonstrate that ferro-orbital
fluctuations enhance the nematic susceptibility, cooperatively promoting
an electronic tetragonal symmetry-breaking. Finally, we argue that
$T_{c}$ in the iron pnictides might be enhanced due to nematic fluctuations
of magnetic origin. 
\end{abstract}
\maketitle

\section{Introduction}

Understanding the normal state properties of the iron-based superconductors
is a key step to decipher the nature of their high-temperature superconducting
state (for reviews, see \cite{reviews}). In fact, the superconducting
(SC) transition temperature $T_{c}$ is maximum near an instability
to a spin-density wave state (SDW), suggesting that spin fluctuations
may play an important role for the formation of the Cooper pairs \cite{magnetic}.
However, besides this SDW state, another ordered state is observed
in the phase diagrams of the iron pnictides: the orthorhombic (Ort)
phase. Unlike other superconductors that also display lattice instabilities
- such as the cuprates - in the pnictides the Ort transition line
always follows the SDW transition line across the phase diagram, even
when the latter is bent-back inside the SC dome \cite{FernandesPRB10,Nandi09}.
Most importantly, the Ort transition at $T_{s}$ either precedes or
is simultaneous to the SDW transition at $T_{N}$, implying that the
lattice distortion is not a mere consequence of long-range magnetic
order. Since these two phases are closely related, and since magnetic
fluctuations are the main candidate to explain the high-$T_{c}$ SC
state, it is natural to investigate in more detail the origin of the
Ort phase and its interplay with SC.

On the experimental side, the development of detwinning techniques
(see \cite{Fisher11} for a review) allowed researchers to investigate
the anisotropic properties of the Ort phase using several different
experimental probes \cite{Chu10,Tanatar10,Davis10,Duzsa11,Uchida11,Shen11,Song11,Matsuda11}.
It became clear that the small lattice distortion could explain neither
the magnitude nor the doping-dependence of these anisotropies - in
particular, the remarkable amplitude of the resistivity anisotropy
in electron-doped samples \cite{Chu10,Kuo11} and its sign-change
upon hole doping \cite{Blomberg12}. Instead, this pool of observations
suggest that the Ort phase is actually a manifestation of another
electronic ordered phase that breaks the same tetragonal symmetry
of the system.

\begin{figure}

\begin{centering}
\includegraphics[width=1\columnwidth]{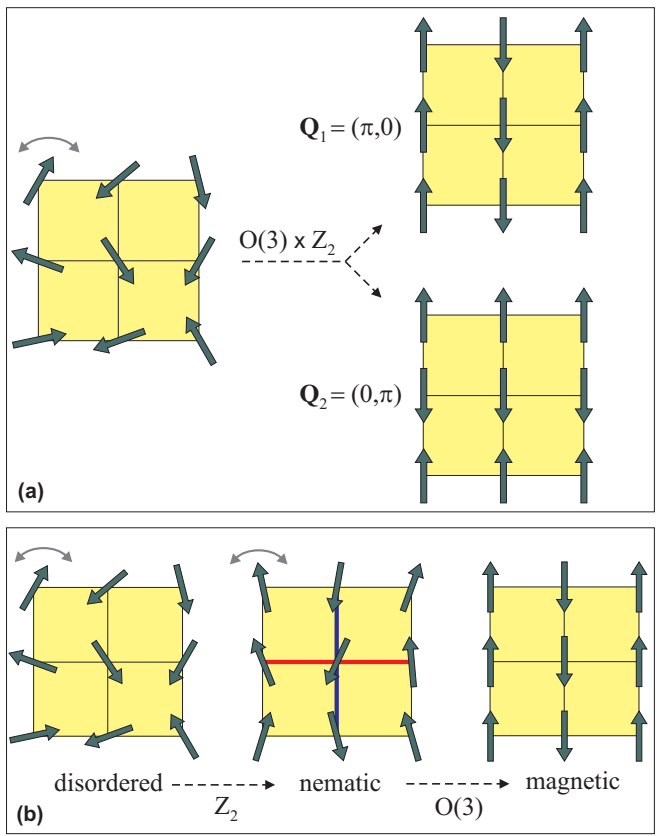} 
\par\end{centering}

\caption{Schematic representation of the nematic transition in real space.
(a) The transition from the disordered phase to the SDW phase breaks
an $O(3)\times Z_{2}$ symmetry. The $O(3)$ symmetry refers to rotations
in spin space while the $Z_{2}$ (Ising) symmetry refers to the two
degenerate ground states of magnetic stripes with parallel spins along
the $y$ axis (ordering vector $\mathbf{Q}_{1}=\left(\pi,0\right)$)
or along the $x$ axis (ordering vector $\mathbf{Q}_{2}=\left(0,\pi\right)$).
(b) The $O(3)\times Z_{2}$ symmetry can be broken in two steps. First,
only the $Z_{2}$ symmetry is broken: the system is still paramagnetic,
since $\left\langle \mathbf{S}_{i}\right\rangle =0$ (indicated by
the gray double-arrow on top of the spins), but the spin correlations
break the tetragonal symmetry, $\left\langle \mathbf{S}_{i}\cdot\mathbf{S}_{i+\mathbf{x}}\right\rangle =-\left\langle \mathbf{S}_{i}\cdot\mathbf{S}_{i+\mathbf{y}}\right\rangle $
(red and blue bonds, respectively). In the second step, the $O\left(3\right)$
symmetry is broken and the system acquires long-range magnetic order. }

\label{fig_nematic_real_space} 
\end{figure}

On the theoretical side, two different approaches have been proposed
to explain the connection between the Ort and SDW phases, as well
as the anisotropic properties displayed below $T_{s}$. In one approach,
the onset of the SDW is regarded as a consequence of the Ort phase,
which itself is driven by a spontaneous ferro-orbital order \cite{kruger09,RRPSingh09,w_ku10,Phillips10,devereaux10,Nevidomskyy11}.
In this scenario, the anisotropic properties displayed in the Ort
phase are a consequence of the unequal occupations between the $d_{xz}$
and $d_{yz}$ orbitals of the Fe atom. Indeed, a spontaneous orbital
polarization is obtained in some (but not all, see Ref. \cite{kruger09})
strong-coupling models - in particular, in some versions of the Kugel-Khomskii
model designed for the iron pnictides. However, in weak-coupling or
moderate-coupling models, orbital order is usually found only inside
the SDW phase, but not as a spontaneous instability of the system
preempting the SDW transition \cite{bascones,dagotto,kotliar}.

The other approach adopts the opposite point of view, i.e. that the
SDW phase causes the Ort phase \cite{Fang08,Xu08,FernandesPRL10,Si08,Qi09,Mazin09,Antropov11,JiangpingHu11,Batista11,Lorenzana_nematic,FernandesPRB12}.
Since $T_{s}\geq T_{N}$, it is not long-range magnetic order, but
magnetic fluctuations, that spontaneously break the tetragonal symmetry
of the system. Interestingly, similar ideas were proposed in the early
1970s to explain the splitting between the magnetic and the cubic-to-tetragonal
transition in a family of rare-earth pnictides, such as DySb, CeSb,
DyP, HoP, DyAs \cite{Levy71}.

The qualitative idea is simple and can be understood using symmetry
arguments, as shown in Fig. \ref{fig_nematic_real_space}. In many
antiferromagnets, the symmetry that is broken at the magnetic transition
temperature is the $O(3)$ spin-rotational symmetry. To the $O(3)$
symmetry breaking corresponds also a translational symmetry breaking,
due to the increase in the size of the crystalline unit cell in the
magnetically ordered phase. In the iron pnictides, however, the situation
is different. The SDW ground state is actually doubly-degenerate,
as it is characterized by magnetic stripes of parallel spins along
either the $y$ axis (ordering vector $\mathbf{Q}_{1}=\left(\pi,0\right)$)
or the $x$ axis (ordering vector $\mathbf{Q}_{2}=\left(0,\pi\right)$).
Therefore, to go to the ordered state, the system has to break not
only the $O(3)$ spin-rotational symmetry, but it also has to choose
between two degenerate ground states, which corresponds to a $Z_{2}$
(Ising-like) symmetry. Since $Z_{2}$ is a discrete symmetry, the
$Z_{2}$ symmetry-breaking is expected to be less affected by fluctuations
than the continuous $O(3)$ symmetry-breaking, what opens up the possibility
of the former happening before the latter.

This is the idea behind the Ising-nematic state: an intermediate phase
preempting the SDW state, where the $Z_{2}$ symmetry is broken but
the $O(3)$ symmetry is not. In real space, the $Z_{2}$ symmetry-breaking
corresponds to a broken tetragonal symmetry, since the correlation
functions $\left\langle \mathbf{S}_{i}\cdot\mathbf{S}_{i+\mathbf{x}}\right\rangle $
and $\left\langle \mathbf{S}_{i}\cdot\mathbf{S}_{i+\mathbf{y}}\right\rangle $
acquire opposite signs (see Fig. \ref{fig_nematic_real_space}(b)).
This is the origin of the term nematic: in liquid crystals, a nematic
phase is characterized by a broken rotational symmetry and an unbroken
translational symmetry. Although the translational and rotational
symmetries are always broken in crystalline solids, the analogy remains
valid: in the electronic nematic phase, the point-group symmetry is
reduced from $C_{4}$\textbf{\ }(tetragonal) to $C_{2}$ (orthorhombic),
corresponding to an additional lowering of the rotational symmetry.
From a purely symmetry point of view, the nematic state is therefore
equivalent to the orthorhombic phase, which is the result of the inevitably
induced distortion of the crystalline lattice (see below). The term
{}``nematic'' is however used to emphasize the fact that the phase
transition is of purely electronic origin and would still take place
in a perfectly rigid lattice (for nematic transitions in other systems,
see Ref. \cite{Fradkin_review}). The recent analysis of Chu \emph{et
al.}\cite{IanLegendre} demonstrated that it is indeed possible to
experimentally determine the driving force behind the transition in
the iron pnictides and distinguish between electronic driven transitions
and ordinary tetragonal-to-orthogonal structural transitions\textbf{.} 

The mechanism behind the spontaneous breaking of the additional Ising
symmetry is reminiscent of the order-out-of-disorder mechanism put
forward by Chandra, Coleman, and Larkin in the context of localized
spin models \cite{chandra}. Not surprisingly, the first model calculations
that obtained an spontaneous nematic phase in the pnictides were based
on a strong-coupling approach, the so-called $J_{1}$-$J_{2}$ model
\cite{Fang08,Xu08,Si08}. More recently, it was shown that an itinerant
description of the system also accounts for a preemptive nematic phase
\cite{FernandesPRB12}. Notwithstanding important differences between
the strong-coupling and weak-coupling approaches - in particular on
the character of the nematic and magnetic transitions as function
of doping and pressure \cite{FernandesPRB12} - they share similar
physics: magnetic fluctuations spontaneously break the tetragonal
symmetry already in the paramagnetic phase.

In this paper, we will focus not on the origin of the nematic degrees
of freedom, but rather on its manifestations in several properties
of the iron pnictides, such as the spin-spin correlation function,
the spin-lattice relaxation rate, the resistivity anisotropy, the
uniform magnetic susceptibility, and the elastic modulus. Since all
these quantities are experimentally accessible, they provide important
cornerstones to test the predictions of the nematic model. In particular,
we will investigate not only the role played by nematic order but
also by nematic fluctuations. The former is trivially manifested as
a finite orthorhombic distortion of the lattice, but, due to its magnetic
origin, it also leaves distinctive signatures in several magnetic
properties. For instance, the onset of nematic order enhances magnetic
fluctuations, causing a kink in the spin-lattice relaxation rate,
and splits the degenerate spectrum of magnetic excitations around
$\mathbf{Q}_{1}=\left(\pi,0\right)$ and $\mathbf{Q}_{2}=\left(0,\pi\right)$.
Furthermore, we will show that the anisotropic uniform susceptibility
and the resistivity anisotropy are proportional to the nematic order
parameter near $T_{s}$, although the sign of the latter depends on
the geometry of the Fermi surface.

\begin{figure}

\begin{centering}
\includegraphics[width=0.85\columnwidth]{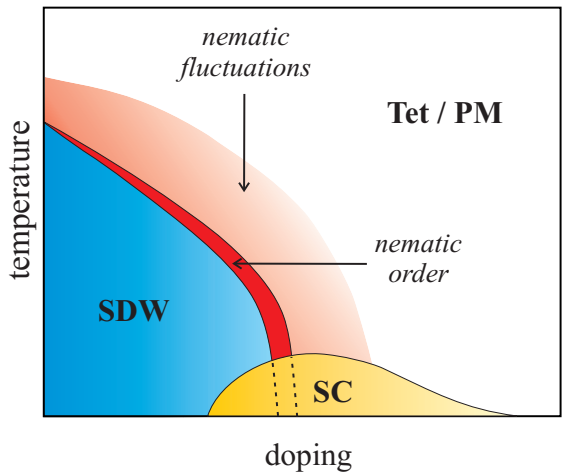} 
\par\end{centering}

\caption{Schematic phase diagram of the iron pnictides, emphasizing the role
played by nematic order and nematic fluctuations. SDW denotes the
spin-density wave state, SC the superconducting phase, PM the paramagnetic
phase, and Tet the tetragonal phase. Tetragonal symmetry is only broken
below the nematic/orthorhombic transition line, but nematic fluctuations
remain at higher temperatures, as evidenced by shear modulus \cite{FernandesPRL10,Yoshizawa12},
uniform susceptibility anisotropy \cite{Matsuda11}, and resistivity
anisotropy \cite{Chu10,Tanatar10} measurements. }

\label{fig_phase_diagrams} 
\end{figure}

Besides nematic order, low-energy nematic fluctuations also emerge,
strongly affecting the system properties. These fluctuations renormalize
the shear modulus already at high temperatures, providing a way of
probing the nematic susceptibility directly from elastic measurements.
We will also discuss the interplay between nematicity and the other
electronic degrees of freedom of the system. In particular, we will
show how ferro-orbital fluctuations can enhance the tendency to a
nematic instability, which in turn can promote orbital order. Finally,
the interaction between nematic and superconducting degrees of freedom
will be addressed. As a consequence of its magnetic origin and of
the competition between SDW and SC, nematicity indirectly competes
with SC. This is manifested in the suppression of the orthorhombic
distortion below $T_{c}$ and in the hardening of the shear modulus
inside the SC dome. Despite this competition, we will give a qualitative
argument of why nematic fluctuations can enhance the value of $T_{c}$
without affecting the symmetry of the SC state. The schematic phase
diagram shown in Fig. \ref{fig_phase_diagrams} summarizes the main
points of this work: not only nematic order, but also nematic fluctuations,
play an important role in the physics of the iron pnictides.

The paper is organized as follows: in Section \ref{sec_model} we
review the model that relates the nematic order parameter and the
nematic susceptibility to the magnetic spectrum of the system. In
Section \ref{sec_manifestations} we use this model to calculate several
system properties that can be probed experimentally: the magnetic
properties in the PM-Ort phase (spin-lattice relaxation rate, spin-spin
correlation function, and uniform susceptibility); the electronic
spectral function in the PM-Ort phase; the orthorhombic distortion
in the PM-Ort phase and the shear modulus in the PM-Tet phase; and
the resistivity anisotropy above the SDW transition temperature. We
address the interplay between nematic and ferro-orbital degrees of
freedom in Section \ref{sec_orbital_order}, while the interplay with
superconductivity is investigated in Section \ref{sec_nematic_SC}.
Section \ref{sec_conclusion} is devoted to the conclusions and closing
remarks.

\section{Magnetic model for the nematic phase \label{sec_model}}

We start with an effective action describing the magnetic degrees
of freedom \cite{Eremin10,FernandesPRB12}. Since there are two magnetic
ground states, we introduce two order parameters $\mathbf{M}_{1}$
and $\mathbf{M}_{2}$ corresponding respectively to stripes with parallel
spins along the $y$ axis (i.e. ordering vector $\mathbf{Q}_{1}=\left(\pi,0\right)$)
and stripes with parallel spins along the $x$ axis (i.e. ordering
vector $\mathbf{Q}_{2}=\left(0,\pi\right)$). Thus, the spatial magnetic
configuration is given by $\mathbf{S}\left(\mathbf{r}\right)=\mathbf{M}_{1}\mathrm{e}^{i\mathbf{Q}_{1}\cdot\mathbf{r}}+\mathbf{M}_{2}\mathrm{e}^{i\mathbf{Q}_{2}\cdot\mathbf{r}}$.
The free energy describing these degrees of freedom is given by:

\begin{eqnarray}
F_{\mathrm{mag}} & = & \int_{q}\chi_{0}^{-1}\left(q\right)\left(\mathbf{M}_{1,q}\cdot\mathbf{M}_{1,-q}+\mathbf{M}_{2,q}\cdot\mathbf{M}_{2,-q}\right)\nonumber \\
 &  & +\frac{u}{2}\int_{x}\left(M_{1}^{2}+M_{2}^{2}\right)^{2}-\frac{g}{2}\int_{x}\left(M_{1}^{2}-M_{2}^{2}\right)^{2}\label{F_mag}\end{eqnarray}
 where $\int_{q}=T\sum_{n}\int\frac{d^{d}q}{\left(2\pi\right)^{d}}$
and $q=\left(\mathbf{q},\omega_{n}\right)$ denotes the momentum $\mathbf{q}$
and the Matsubara frequency $\omega_{n}=2n\pi T$. Here $\chi_{0}^{-1}\left(q\right)=r_{0}+q^{2}+\gamma^{-1}\left|\omega_{n}\right|$
is the bare dynamic spin susceptibility and $r_{0}\propto\left(T-T_{N,0}\right)$,
with $T_{N,0}$ the mean-field SDW transition temperature and $\gamma$
the Landau damping. Notice that $\mathbf{M}_{i,q}$ is a function
of both momentum and Matsubara frequency, as it has both static and
dynamic fluctuations.

The coupling constants $u$ and $g$ determine the nature of the magnetic
ground state. Consider, for instance, the uniform limit and minimize
the free energy $F$ with respect to $M_{i}$. Then, if $g>0$ and
$u>0$, we obtain two solutions for $T<T_{N,0}$ corresponding to
$M_{2}=0$ and $M_{1}=\sqrt{-r_{0}/u}$ or $M_{1}=0$ and $M_{2}=\sqrt{-r_{0}/u}$.
These are nothing but the two degenerate magnetic stripe configurations
we discussed earlier. Therefore, hereafter we consider $u>0$ and
$g>0$. From symmetry considerations, the action (\ref{F_mag}) can
also contain the term $\left(\mathbf{M}_{1}\cdot\mathbf{M}_{2}\right)^{2}$,
which, for $g>0$, does not affect the selection of the ground state
as long as its coefficient is not a large negative number compared
to $g$.

Although here we introduced the free energy (\ref{F_mag}) in a phenomenological
way, it can be derived microscopically from both the itinerant \cite{FernandesPRB12}
and localized-spin approaches \cite{Fang08}. In both cases, the term
$\left(\mathbf{M}_{1}\cdot\mathbf{M}_{2}\right)^{2}$ is absent. Even
though both approaches find $g>0$, which selects the magnetic ground
states observed in the iron pnictides, the magnitude of this coupling
constant is different in the band model and in the $J_{1}-J_{2}$
model. In particular, while in the latter $g$ is small and induced
by thermal or quantum fluctuations, in the former it can be large,
as it is given by products of the non-interacting electronics Green's
functions. It has been suggested that a relatively large $g$ is necessary
in order to fit the spin-wave spectrum inside the SDW phase \cite{Antropov11}.

In terms of the magnetic degrees of freedom, the nematic order parameter
can be expressed as $\varphi\propto M_{1}^{2}-M_{2}^{2}$. Indeed,
$\left\langle \varphi\right\rangle \neq0$ implies that the fluctuations
around one of the ordering vectors $\left\langle M_{1}^{2}\right\rangle $
are on average different than the fluctuations around the other ordering
vector $\left\langle M_{2}^{2}\right\rangle $. Since $\mathbf{Q}_{1}=\left(\pi,0\right)$
and $\mathbf{Q}_{2}=\left(0,\pi\right)$ are related by a $90^{\circ}$
rotation in the square-lattice unit cell, $\left\langle M_{1}^{2}\right\rangle \neq\left\langle M_{2}^{2}\right\rangle $
implies that the $x$ and $y$ directions are inequivalent and the
tetragonal symmetry is broken. This tetragonal symmetry-breaking is
enforced by magnetic fluctuations, and not long-range magnetic order,
since $\left\langle M_{1}^{2}\right\rangle \neq\left\langle M_{2}^{2}\right\rangle $
does not require or implies $\left\langle \mathbf{M}_{i}\right\rangle =0$.

Once long-range nematic order sets in, it has a feedback effect on
the magnetic fluctuations, assisting the transition to the SDW state.
Physically, the nematic order parameter {}``transfers'' magnetic
spectral weight from one ordering vector to the other, enhancing magnetic
fluctuations in one channel but suppressing them on the other (see
Section \ref{sec_manifestations}C). This process increases the SDW
transition temperature, as it becomes larger than what it would be
in the absence of long-range nematic order.

The expressions relating the nematic order parameter $\varphi$ and
the magnetic susceptibility can be obtained by introducing two auxiliary
Hubbard-Stratonovich fields in the free energy (\ref{F_mag}), one
for each quartic term (for more details, see \cite{FernandesPRB12}).
The first one accounts for the Gaussian fluctuations of the magnetic
order parameter, $\left\langle M_{1}^{2}+M_{2}^{2}\right\rangle $,
whereas the second one accounts for the possibility of a finite nematic
order parameter, $\left\langle M_{1}^{2}-M_{2}^{2}\right\rangle $.
In the saddle-point ($1/N$) approximation \cite{comment}, one obtains
a set of non-linear coupled equations for these two auxiliary fields:

\begin{eqnarray}
\varphi & = & \frac{g}{2}\int_{q}\left[\tilde{\chi}_{1}\left(q\right)-\tilde{\chi}_{2}\left(q\right)\right]\nonumber \\
r & = & r_{0}+\frac{u}{2}\int_{q}\left[\tilde{\chi}_{1}\left(q\right)+\tilde{\chi}_{2}\left(q\right)\right]\label{self_cons_eq}\end{eqnarray}
 where $r\propto\xi^{-2}$, with $\xi$ denoting the magnetic correlation
length renormalized by Gaussian fluctuations. $\tilde{\chi}_{i}\left(q\right)$
is the renormalized magnetic susceptibility at the ordering vector
$\mathbf{Q}_{i}$:

\begin{eqnarray}
\tilde{\chi}_{1}\left(q\right) & = & \frac{1}{\left(r-\varphi\right)+q^{2}+\gamma^{-1}\left|\omega_{n}\right|}\nonumber \\
\tilde{\chi}_{2}\left(q\right) & = & \frac{1}{\left(r+\varphi\right)+q^{2}+\gamma^{-1}\left|\omega_{n}\right|}\label{magnetic_suscept}\end{eqnarray}

Expanding equations (\ref{self_cons_eq}) for small $\varphi$, we
obtain a familiar $\phi^{4}$-type equation of state for $\varphi$:

\begin{equation}
\left(1-g\int_{q}\chi^{2}\left(q\right)\right)\varphi+b\varphi^{3}+\mathcal{O}\left(\varphi^{5}\right)=0\label{equation_state}\end{equation}
 where $\chi\left(q\right)$ has the same form as the bare susceptibility
$\chi_{0}\left(q\right)$ but with $r_{0}$ replaced by $r$, i.e.
$\chi^{-1}\left(q\right)=r+q^{2}+\gamma^{-1}\left|\omega_{n}\right|$.
The sign of the cubic coefficient $b$ depends on the ratio $u/g$
and on the dimensionality $d$ \cite{FernandesPRB12}, and determines
whether the nematic transition is first-order or second-order.

For $b>0$, the sign of the linear coefficient determines when a finite
$\varphi\neq0$ becomes a solution of the equation of state (\ref{equation_state}).
At high enough temperatures, the magnetic fluctuations are weak and
$\int_{q}\chi^{2}\left(q\right)\propto\xi^{4-d}$ is small, implying
that the linear coefficient is positive. Near a magnetic instability,
however, $\int_{q}\chi^{2}\left(q\right)$ increases significantly,
since it diverges at the magnetic transition temperature for $d+z\leq4$
($z$ is the dynamic critical exponent, relevant for quantum phase
transitions). Therefore, proximity to an SDW transition makes the
linear coefficient change sign, inducing the nematic phase transition.
Notice that, for $b<0$, the first-order transition may take place
while the linear coefficient is still positive. 

Another way of expressing the same results is via the nematic susceptibility,
defined as the correlation function $\chi_{\mathrm{nem}}\left(q\right)=\left\langle \left(M_{1}^{2}-M_{2}^{2}\right)_{q}\left(M_{1}^{2}-M_{2}^{2}\right)_{-q}\right\rangle $.
After introducing a symmetry-breaking term $h\left(M_{1}^{2}-M_{2}^{2}\right)$
in the free energy (\ref{F_mag}) and then taking $h\rightarrow0$,
we obtain:

\begin{equation}
\chi_{\mathrm{nem}}=\frac{1}{g}\left(\frac{\left\langle \varphi^{2}\right\rangle }{g}-1\right)\label{aux_nematic_suscept}\end{equation}

In the saddle-point approximation, it follows that \cite{FernandesPRL10}:

\begin{equation}
\chi_{\mathrm{nem}}=\frac{\int_{q}\chi^{2}\left(q\right)}{1-g\int_{q}\chi^{2}\left(q\right)}\label{nematic_suscept}\end{equation}

As expected, the condition for the onset of long-range magnetic order,
$\chi_{\mathrm{nem}}\rightarrow\infty$, agrees with the condition
for a finite $\varphi\neq0$ in the equation of state (\ref{equation_state}).

\begin{figure}
\begin{centering}
\includegraphics[width=1\columnwidth]{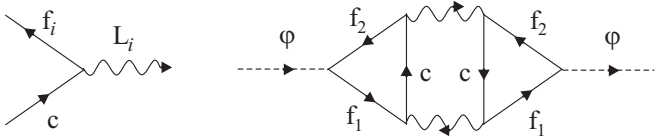}
\par\end{centering}

\caption{Diagrammatic representation of the nematic susceptibility in terms
of the electronic operators $c_{\mathbf{k}}$, $f_{1,\mathbf{k}}=c_{\mathbf{k}+\mathbf{Q}_{1}}+c_{\mathbf{k}+\mathbf{Q}_{2}}$
and $f_{2,\mathbf{k}}=c_{\mathbf{k}+\mathbf{Q}_{1}}-c_{\mathbf{k}+\mathbf{Q}_{2}}$,
where $c_{\mathbf{k}}$ is the operator associated with states around
the center of the Brillouin zone and $\mathbf{Q}_{i}$ are the magnetic
ordering vectors. Dashed lines are the external nematic legs, solid
lines are the electronic propagators, and the wavy lines are the magnetic
propagators associated with $\mathbf{L}_{1}=\mathbf{M}_{1}+\mathbf{M}_{2}$
and $\mathbf{L}_{2}=\mathbf{M}_{1}-\mathbf{M}_{2}$. The nematic susceptibility
(on the right) corresponds to an Aslamazov-Larkin type diagram. \label{fig_aslamazov_larkin}}

\end{figure}

It is interesting to obtain a more microscopic understanding of the
nematic susceptibility in terms of the electrons. For this, we consider
the states around the center of the square-lattice Brillouin zone
(annihilation operator $c_{\mathbf{k}}$), where the hole pockets
are present, and the states $c_{\mathbf{k}+\mathbf{Q}_{i}}$ in the
electron pockets centered at the the magnetic ordering vectors $\mathbf{Q}_{1}=\left(\pi,0\right)$
and $\mathbf{Q}_{2}=\left(0,\pi\right)$. The basic nematic vertex
is given by $\varphi\left(M_{1}^{2}-M_{2}^{2}\right)$, and the collective
spin modes are expressed in terms of the electronic operators as $\mathbf{M}_{i}=\sum_{\mathbf{k}}c_{\mathbf{k}\alpha}^{\dagger}\boldsymbol{\sigma}_{\alpha\beta}c_{\mathbf{k}+\mathbf{Q}_{i}\beta}$,
with Pauli matrices $\boldsymbol{\sigma}_{\alpha\beta}$. It is convenient
to introduce the two-sublattice Neel vectors $\mathbf{L}_{1}$ and
$\mathbf{L}_{2}$, defined as $\mathbf{L}_{1}=\mathbf{M}_{1}+\mathbf{M}_{2}$
and $\mathbf{L}_{2}=\mathbf{M}_{1}-\mathbf{M}_{2}$. Then, it follows
that $\mathbf{L}_{i}=\sum_{\mathbf{k}}c_{\mathbf{k}\alpha}^{\dagger}\boldsymbol{\sigma}_{\alpha\beta}f_{i,\mathbf{k}\beta}$,
where we introduced the fermionic operators $f_{1,\mathbf{k}}=c_{\mathbf{k}+\mathbf{Q}_{1}}+c_{\mathbf{k}+\mathbf{Q}_{2}}$
and $f_{2,\mathbf{k}}=c_{\mathbf{k}+\mathbf{Q}_{1}}-c_{\mathbf{k}+\mathbf{Q}_{2}}$.
Within this notation, the nematic vertex becomes $\varphi\left(\mathbf{L}_{1}\cdot\mathbf{L}_{2}\right)$,
and the nematic susceptibility is given by an Aslamazov-Larkin type
diagram, as shown schematically in Fig. \ref{fig_aslamazov_larkin}.

Notice that $\chi_{\mathrm{nem}}$ is a four-spin correlation function.
In a regular antiferromagnet, where only one order parameter is present,
its critical behavior can be fully expressed in terms of the critical
behavior of the two-spin correlation function - the magnetic susceptibility
$\chi\left(q\right)$. In that case, the four-spin correlation function
does not contain any information that is not already embedded in the
two-spin function. In the present case, however, this is not true:
due to the nematic degrees of freedom, the four-spin correlation function
acquires its own critical behavior, revealing the emergent character
of the nematic phase.

In our subsequent analysis we primarily focus on the anomalies that
take place at the nematic transition. For a realistic description
of the pnictides the behavior at the nearby magnetic transition is
of course closely connected to that at the Neel temperature (see Fig.
\ref{fig_kink}). However, in order to stress the behavior of the
paramagnetic nematic phase we will frequently consider either two-dimensional
systems, where the magnetic ordering temperature is suppressed to
zero, or systems where the splitting between the two phase transitions
is not too small.

\section{Manifestations of nematic order and nematic fluctuations in the normal
state \label{sec_manifestations}}

One immediate consequence of the coupled non-linear equations (\ref{self_cons_eq})
is that the nematic and magnetic transitions tend to follow each other.
For instance, the divergence of the static nematic susceptibility
in Eq. (\ref{nematic_suscept}) depends on how close the system is
to a magnetic instability, where $\int_{q}\chi^{2}\left(q\right)$
diverges. At the same time, once long-range nematic order sets in,
the static magnetic susceptibility $\tilde{\chi}_{i}\left(q=0\right)$
is enhanced, as shown in Eq. (\ref{magnetic_suscept}), bringing the
SDW transition temperature up. Therefore, both the nematic and the
magnetic transitions are tied together, in qualitative agreement with
the experimental phase diagrams of the iron pnictides. This is perhaps
the most striking feature in favor of the nematic model.

The behavior of the solutions of Eq. (\ref{self_cons_eq}) has been
studied in detail elsewhere, and the characters of the magnetic and
nematic transitions have been investigated for a wide range of parameters
\cite{FernandesPRB12}. In the remainder of the paper, we will focus
instead on the manifestations of long-range nematic order, as well
as nematic fluctuations, in several properties of the system.

\subsection{Orthorhombic distortion: x-ray diffraction}

In the nematic model, the spontaneous breaking of the tetragonal symmetry
takes place in the magnetic sector, since magnetic fluctuations become
stronger around one of the two ordering vectors $\mathbf{Q}_{1}=\left(\pi,0\right)$
and $\mathbf{Q}_{2}=\left(0,\pi\right)$, which are related by a $90^{\circ}$
rotation. Following Landau's theory of phase transitions, once the
tetragonal symmetry is broken in the magnetic sector, it will be broken
in all sectors at the same temperature. Indeed, if one constructs
another (scalar) order parameter $\eta$ that breaks tetragonal symmetry,
it must couple bi-linearly to the nematic order parameter $\varphi$
in the free-energy expansion, implying that $\eta\propto\varphi$
in the free-energy minimum.

This is the case for the orthorhombic order parameter $\varepsilon_{s}\equiv\left(a-b\right)/\left(a+b\right)$,
where $a$ and $b$ are the lattice constants. Considering a harmonic
lattice, the elastic free energy is given by:

\begin{equation}
F_{\mathrm{el}}=\int_{x}\frac{C_{s}}{2}\,\varepsilon_{s}^{2}-\lambda_{\mathrm{el}}\int_{x}\varepsilon_{s}\left(M_{1}^{2}-M_{2}^{2}\right)\label{F_el}\end{equation}
 where $C_{s}$ is the shear modulus and $\lambda_{\mathrm{el}}$
is the magneto-elastic coupling. If one works in the coordinate system
of the square lattice with one Fe per unit cell, then $C_{s}\equiv C_{11}-C_{12}$;
instead, if one works in the coordinate system with two Fe per unit
cell, then $C_{s}\equiv C_{66}$. In either case, minimization of
Eq. (\ref{F_el}) leads to $\varepsilon_{s}\propto\left\langle M_{1}^{2}\right\rangle -\left\langle M_{2}^{2}\right\rangle $,
i.e. measuring the orthorhombic distortion is equivalent to measuring
the nematic order parameter. Of course, the mere detection of an orthorhombic
distortion is by no means an evidence for the electronic character
of the structural transition. Yet, one can ask whether there are any
signatures of the magnetic character of the structural transition
in the behavior of $\varepsilon_{s}$.

\begin{figure}

\begin{centering}
\includegraphics[width=1\columnwidth]{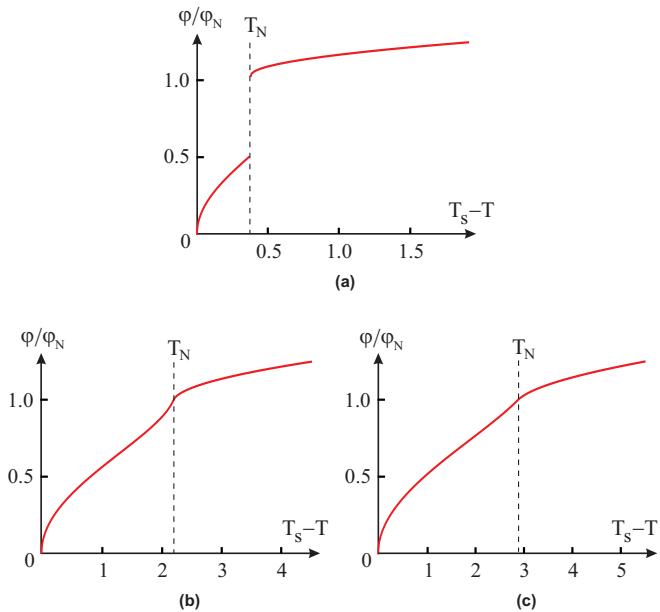} 
\par\end{centering}

\caption{Nematic order parameter $\varphi$ (in units of its value $\varphi_{N}$
at the magnetic transition temperature $T_{N}$) as function of the
reduced temperature $T_{s}-T$ (arbitrary units). In panel (a), the
split magnetic transition is first-order, while in (b) and (c) it
is second-order and progressively farther from the magnetic tricritical
point. $\varphi$ is obtained by solving Eqs. (\ref{self_cons_eq})
for a three-dimensional anisotropic magnetic dispersion. The only
parameter changed in each plot is the ratio $u/g$. }

\label{fig_kink} 
\end{figure}

Interestingly, the solution of the set of self-consistent equations
(\ref{self_cons_eq}) shows that, when the magnetic and nematic transitions
are split, $\varepsilon_{s}$ is in general not affected by the magnetic
transition, except near a magnetic tricritical point of the phase
diagram \cite{FernandesPRB12}. In this region of parameter-space,
$\varepsilon_{s}$ has a kink at the magnetic transition temperature
- more specifically, $\left|d\varepsilon_{s}/dT\right|_{T_{N}^{-}}<\left|d\varepsilon_{s}/dT\right|_{T_{N}^{+}}$,
as shown in Fig. \ref{fig_kink}. Away from the tricritical point,
however, $\varepsilon_{s}$ becomes a smoother function around $T_{N}$.
Indeed, x-ray diffraction measurements in $\mathrm{Ba\left(Fe_{1-x}Co_{x}\right)_{2}As_{2}}$
find a kink in the orthorhombic distortion only in the proximity of
$x\approx0.02$ \cite{Kim11}, where a magnetic tricritical point
has been experimentally established \cite{Birgeneau11}.

Notice also that the magneto-elastic coupling has important implications
for detwinned samples. Detwinning can be achieved by applying a strain
$\varepsilon_{s}$, which acts as an external field to the nematic
order parameter $\left\langle M_{1}^{2}\right\rangle -\left\langle M_{2}^{2}\right\rangle $
in Eq. (\ref{F_el}), see Ref. \cite{IanLegendre}. Then, one has
to add to the elastic free energy the term $\sigma\varepsilon_{s}$:

\begin{equation}
F_{\mathrm{el}}=\int_{x}\frac{C_{s}}{2}\,\varepsilon_{s}^{2}-\lambda_{\mathrm{el}}\int_{x}\varepsilon_{s}\left(M_{1}^{2}-M_{2}^{2}\right)-\int_{x}\sigma\varepsilon_{s}\label{F_el_stress}\end{equation}

For a harmonic lattice, the elastic contribution to the partition
function $Z=\int dM_{i}\, d\varepsilon_{s}\mathrm{e}^{-F_{\mathrm{mag}}-F_{\mathrm{el}}}$
can be integrated out analytically, yielding the renormalized magnetic
free energy:

\begin{eqnarray}
\tilde{F}_{\mathrm{mag}} & = & F_{\mathrm{mag}}\left[\mathbf{M}_{1},\mathbf{M}_{2}\right]-\frac{\lambda_{\mathrm{el}}^{2}}{2C_{s}}\int_{x}\left(M_{1}^{2}-M_{2}^{2}\right)^{2}\nonumber \\
 &  & -\frac{\sigma^{2}}{2C_{s}}-\frac{\sigma\lambda_{\mathrm{el}}}{C_{s}}\int_{x}\left(M_{1}^{2}-M_{2}^{2}\right)\label{F_mag_el_renormalized}\end{eqnarray}

Comparing to the original action, Eq. (\ref{F_mag}), the nematic
coupling $g$ is enhanced \cite{FernandesPRL10} (which holds also
for twin samples) and $\sigma$ is converted into an external field
for the nematic order parameter, whose amplitude depends on $\lambda_{\mathrm{el}}/C_{s}$.
The impact of this external field to the nematic and magnetic transition
temperatures has been widely discussed both theoretically \cite{Hu12,Cano12}
and experimentally \cite{IanLegendre,Blomberg11,Dhital12}.

\subsection{NMR $1/T_{1}T$ spin-lattice relaxation rate}

In the nematic model, magnetic fluctuations are responsible for the
structural transition. Thus, it is natural to expect that the magnetic
properties of the system will undergo noticeable changes at the structural
(nematic) transition temperature $T_{s}$. We first study the $1/T_{1}T$
spin-lattice relaxation rate, which can be directly probed by NMR:

\begin{equation}
\frac{1}{T_{1}T}=\gamma_{\mathrm{g}}^{2}\lim_{\omega\rightarrow0}\sum_{\mathbf{q}}A^{2}\left(\mathbf{q}\right)\frac{\mathrm{Im}\left[\chi_{\mathrm{mag}}\left(\mathbf{q},\omega\right)\right]}{\omega}\label{T1T}\end{equation}

\begin{figure}

\begin{centering}
\includegraphics[width=0.9\columnwidth]{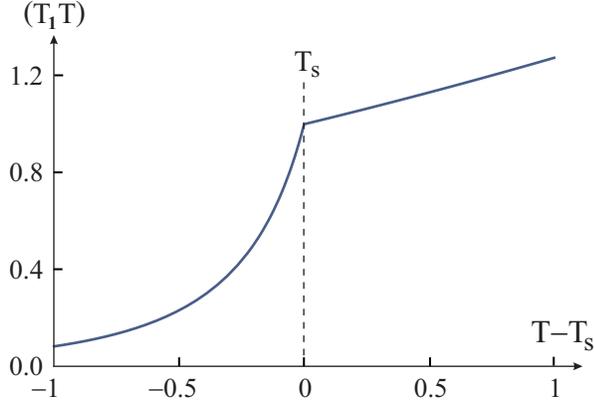} 
\par\end{centering}

\caption{Inverse of the spin-lattice relaxation rate $T_{1}T$ as function
of the reduced temperature $T-T_{s}$ for a two-dimensional system.
The change in the slope at $T_{s}$ is evident, and should be present
whenever the magnetic and structural transitions are split. }

\label{fig_NMR} 
\end{figure}

Here $\gamma_{\mathrm{g}}$ is the gyromagnetic ratio and $A\left(\mathbf{q}\right)$
is the structure factor of the hyperfine interaction. In our system,
$\chi_{\mathrm{mag}}\left(\mathbf{q},\omega\right)$ is strongly peaked
at the ordering vectors $\mathbf{Q}_{1}=\left(\pi,0\right)$ and $\mathbf{Q}_{2}=\left(0,\pi\right)$,
with $\chi_{\mathrm{mag}}\left(\mathbf{q}+\mathbf{Q}_{i},\omega\right)=\tilde{\chi}_{i}\left(\mathbf{q},\omega_{n}\rightarrow-i\omega+0^{+}\right)$
given by Eqs. (\ref{aux_nematic_suscept}). Assuming $A\left(\mathbf{q}\right)$
constant, and considering a two-dimensional system for simplicity,
we can evaluate the momentum sum around each ordering vector and obtain:

\begin{equation}
\frac{1}{T_{1}T}=\kappa\left(\frac{r}{r^{2}-\varphi^{2}}\right)\label{T1T_exp}\end{equation}
 where $\kappa>0$ is a constant. In the absence of nematic order,
$\varphi=0$, one recovers the usual result $1/T_{1}T\propto1/\left(T-T_{N}\right)$.
However, if the paramagnetic phase undergoes a nematic transition,
the spin-lattice relaxation rate diverges when $r=\left|\varphi\right|$,
which is the condition for the magnetic transition to take place,
see Eq. (\ref{aux_nematic_suscept}).

Most interestingly, when the nematic/structural and magnetic transitions
are split, $1/T_{1}T$ displays a kink at the nematic transition,
since:

\begin{equation}
\left(\frac{1}{T_{1}T}\right)_{T_{s}^{-}}-\left(\frac{1}{T_{1}T}\right)_{T_{s}^{+}}\propto\frac{\varphi^{2}}{r^{2}}\label{kink_T1T}\end{equation}

In Fig. \ref{fig_NMR}, we illustrate this behavior by presenting
the temperature dependence of $T_{1}T$ as calculated from the solution
of Eq. (\ref{self_cons_eq}) at $d=2$. The change of slope at $T_{s}$
is evident. Experimentally, this feature should be more pronounced
and easier to observe in compounds with well separated magnetic and
structural transitions. Indeed, in NaFeAs, which has $T_{N}\approx40$
K and $T_{s}\approx60$ K, NMR measurements \cite{Ma_NaFeAs,kitagawa_NaFeAs}
have found a very clear change in the slope of $1/T_{1}T$ at $T_{s}$,
in agreement with the predictions of the nematic model.

\subsection{Spin-spin correlation function: neutron scattering}

A more direct probe of the impact of long-range nematic order on the
magnetic spectrum can be given by inelastic neutron scattering, as
it directly measures the spin-spin correlation function $\mathcal{S}\left(\mathbf{q},\omega\right)=\mathrm{Im}\left[\chi_{\mathrm{mag}}\left(\mathbf{q},\omega\right)\right]/\omega$
at the two ordering vectors $\mathbf{Q}_{1}=\left(\pi,0\right)$ and
$\mathbf{Q}_{2}=\left(0,\pi\right)$. In the paramagnetic phase, we
have the usual expressions for overdamped spin excitations:

\begin{eqnarray}
\mathcal{S}\left(\mathbf{q}+\mathbf{Q}_{1},\omega\right) & =\nonumber \\
\frac{\mathcal{S}_{0}}{\left[\left(r-\varphi\right)+\left(1+\eta\right)q_{x}^{2}+\left(1-\eta\right)q_{y}^{2}\right]^{2}+\gamma^{-2}\omega^{2}}\label{spin_spin1}\end{eqnarray}
 and:

\begin{eqnarray}
\mathcal{S}\left(\mathbf{q}+\mathbf{Q}_{2},\omega\right) & =\nonumber \\
\frac{\mathcal{S}_{0}}{\left[\left(r+\varphi\right)+\left(1-\eta\right)q_{x}^{2}+\left(1+\eta\right)q_{y}^{2}\right]^{2}+\gamma^{-2}\omega^{2}}\label{spin_spin2}\end{eqnarray}
 with $r\propto\xi^{-2}$ and $\varphi$ solutions of the self-consistent
equations (\ref{self_cons_eq}) and $-1<\eta<1$. These expressions
were derived from Eq. (\ref{magnetic_suscept}) with the isotropic
magnetic dispersion $q^{2}$ replaced by the anisotropic dispersion
$\left(1\pm\eta\right)q_{x}^{2}+\left(1\mp\eta\right)q_{y}^{2}$ ,
to make comparison with experiments more realistic \cite{Diallo10,Li10}.

\begin{figure}

\begin{centering}
\includegraphics[width=1\columnwidth]{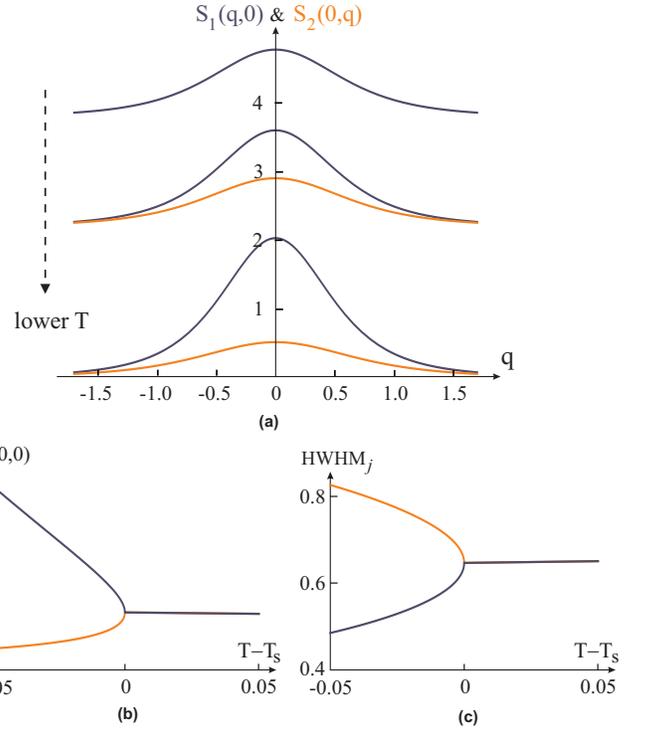} 
\par\end{centering}

\caption{(a) Spin-spin correlation functions $S_{1}\left(q,0\right)\equiv\mathcal{S}\left(q_{x}+\mathbf{Q}_{1},\omega\right)$
(blue) and $S_{2}\left(0,q\right)\equiv\mathcal{S}\left(q_{y}+\mathbf{Q}_{2},\omega\right)$
(orange) as function of momentum transfer $q$ for fixed energy $\omega/\gamma=0.1$
and several temperatures. The upper pairs of curves (coincident) are
for a temperature immediately above the structural transition $T_{s}$
while the lower pairs of curves refer to temperatures progressively
smaller than $T_{s}$. Here the calculations were done at $d=2$.
In (b) and (c), as function of the reduced temperature $T-T_{s}$,
we show the $q=0$ peaks and the half-width at half-maximum HWHM of
each correlation function $S_{1}\left(q,0\right)$ (blue) and $S_{2}\left(0,q\right)$
(orange). }

\label{fig_neutron} 
\end{figure}

The nematic order parameter can be obtained by comparing the correlation
functions $\mathcal{S}\left(q_{x}+\mathbf{Q}_{1},\omega\right)$ and
$\mathcal{S}\left(q_{y}+\mathbf{Q}_{2},\omega\right)$ below $T_{s}$
but still above $T_{N}$ (i.e. in the paramagnetic phase). For a fixed
low energy $\omega/\gamma\ll r$, they have peaks at $q=0$ proportional
to $\left(r\pm\varphi\right)^{-1}$ and half-widths proportional to
$\left(r\pm\varphi\right)^{-1/2}$. In Fig. \ref{fig_neutron}, we
plot the temperature evolution of both $\mathcal{S}\left(q_{x}+\mathbf{Q}_{1},\omega\right)$
and $\mathcal{S}\left(q_{y}+\mathbf{Q}_{2},\omega\right)$, as well
as the temperature dependence of their peaks and widths. We used the
solutions of the self-consistent equations (\ref{self_cons_eq}) for
$d=2$, but the features at $T_{s}$ are of course much more general.

Physically, the onset of long-range nematic order in the paramagnetic
phase enhances the magnetic fluctuations around one of the two ordering
vectors (in this example, $\mathbf{Q}_{1}$) while suppressing the
fluctuations around the other ordering vector ($\mathbf{Q}_{2}$).
As a result, the constant-energy cuts of $\mathcal{S}\left(\mathbf{q}+\mathbf{Q}_{1},\omega\right)$
become narrower and stronger, while the constant-energy cuts of $\mathcal{S}\left(\mathbf{q}+\mathbf{Q}_{2},\omega\right)$
become broader and weaker.

A neutron scattering experiment that detects this behavior will provide
perhaps the most direct evidence for the magnetic origin of the tetragonal
symmetry-breaking in the iron pnictides. Of course, one needs a single
crystal that displays split magnetic and structural transitions, such
as the 1111 and the 111 compounds, or even some underdoped 122 compounds.
The main experimental difficulty is that the crystal needs to be detwinned
- i.e. it must have only one structural domain. With the improvement
of detwinning techniques, it is reasonable to expect that such an
experimental set-up will be available soon.

\subsection{Pseudogap in the spectral function: ARPES}

In the previous two subsections, we discussed the effects of nematic
order on the spectrum of magnetic excitations, and how these effects
can be directly probed via magnetic measurements. Another alternative
is to probe them indirectly via the electronic spectral function,
$A\left(\mathbf{q},\omega\right)=-2\mathrm{Im}\left[G\left(\mathbf{q},\omega\right)\right]$,
where $G\left(\mathbf{q},\omega\right)$ is the electronic Green's
function.

Below $T_{N}$, the electronic band structure is reconstructed by
long-range magnetic order, and $A\left(\mathbf{q},\omega\right)$
is modified due to the folding of shadow bands and the opening of
a SDW gap. Above $T_{N}$, in the paramagnetic phase, the electronic
spectral function is affected by magnetic fluctuations. In particular,
several works have shown that magnetic precursors are able to reduce
the spectral weight of the hot spots at the Fermi level \cite{magn_pseudogap}.
The position $\mathbf{q}_{hs}$ of these hot spots are given by the
relation $E\left(\mathbf{q}_{hs}\right)=E\left(\mathbf{q}_{hs}+\mathbf{Q}_{i}\right)$,
where $E\left(\mathbf{q}\right)$ denotes the band dispersion. To
find the hot spots geometrically, one simply makes a copy of the Fermi
surface and displaces it by the magnetic ordering vector $\mathbf{Q}_{i}$:
the points that overlap with the original Fermi surface correspond
to the hot spots.

\begin{figure}

\begin{centering}
\includegraphics[width=1\columnwidth]{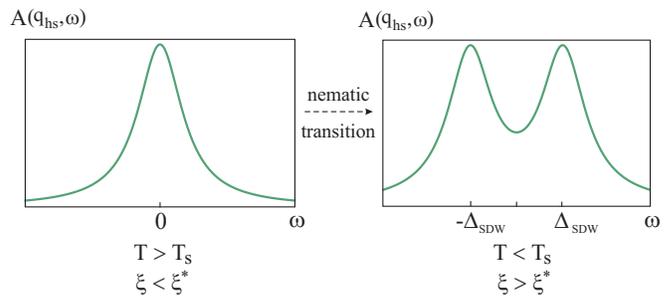} 
\par\end{centering}

\caption{Schematic representation of the opening of a pseudogap in the spectral
function $A\left(\mathbf{q}_{hs},\omega\right)$ at the hot spot $\mathbf{q}_{hs}$
due to long-range nematic order. At high temperatures, above the nematic
transition temperature $T_{s}$, the magnetic correlation length is
below a threshold $\xi^{*}$ and the peak of $A\left(\mathbf{q}_{hs},\omega\right)$
is at $\omega=0$. The onset of nematic order at $T_{s}$ sharply
enhances the magnetic fluctuations, as it was shown in Fig. \protect\ref{fig_NMR},
and $\xi$ can overcome $\xi^{*}$. In this case, the maximum of $A\left(\mathbf{q}_{hs},\omega\right)$
shifts to $\omega\approx\Delta_{\mathrm{SDW}}$, which is the SDW
gap at $T=0$, and $\omega=0$ becomes a minimum. }

\label{fig_pseudogap} 
\end{figure}

Deep inside the magnetically ordered phase, the spectral function
at the hot spots vanishes at the Fermi level, i.e. $A\left(\mathbf{q}_{hs},\omega=0\right)=0$.
At very high temperatures, $A\left(\mathbf{q}_{hs},\omega\right)$
is maximum at $\omega=0$. At intermediate temperatures in the paramagnetic
phase, although $A\left(\mathbf{q}_{hs},\omega=0\right)\neq0$ always,
strong magnetic fluctuations are able to suppress the zero-energy
spectral weight at $\mathbf{q}_{hs}$. Above a certain threshold of
the magnetic correlation length $\xi$, the maximum at $A\left(\mathbf{q}_{hs},\omega=0\right)$
becomes a minimum as a result of spectral weight being transferred
to energies comparable to the zero-temperature SDW gap $\Delta_{\mathrm{SDW}}$,
i.e. the maximum is displaced to $A\left(\mathbf{q}_{hs},\omega=\Delta_{\mathrm{SDW}}\right)$,
see Fig. \ref{fig_pseudogap}. Consequently, the electronic spectrum
has a pseudogap due to the presence of strong magnetic precursors.

We can now understand how nematic order can affect the electronic
spectrum. By sharply enhancing the magnetic fluctuations, as it was
shown before in Fig. \ref{fig_NMR}, the onset of long-range nematic
order may trigger the opening of a pseudogap at the hot spots of the
Fermi surface \cite{FernandesPRB12}. In fact, several experiments
on the iron pnictides have reported possible manifestations of a pseudogap
behavior \cite{Tanatar_pseudogap,Xu_pseudogap}, although more work
is still necessary to pinpoint its exact temperature scale and its
relation to the magnetic and superconducting energy scales. The mechanism
proposed here for the pseudogap could be directly probed by ARPES
or STM measurements, which are sensitive to the electronic spectral
function. Probably the most promising material is NaFeAs, since $T_{s}$
and $T_{N}$ are already below $100$ K and separated by approximately
$20$ K.

\subsection{Uniform susceptibility: torque magnetometry}

Besides affecting the magnetic fluctuations at $\mathbf{Q}_{1}=\left(\pi,0\right)$
and $\mathbf{Q}_{2}=\left(0,\pi\right)$, long-range nematic order
also triggers an anisotropy in the uniform magnetic susceptibility
$\chi_{\mathrm{mag}}\left(\mathbf{q}=0\right)$. To calculate it,
we include the contribution of a non-zero magnetic field $\mathbf{H}$
to the free energy (\ref{F_mag}) and compare the effects of a field
applied along the $x$ direction, $H_{x}$, and a field applied along
the $y$ direction, $H_{y}$. We can either use symmetry considerations
or perform an explicit calculation by adding the Zeeman term to the
electronic dispersions. Neglecting the cross terms $H_{x}H_{y}$,
we obtain the general form:

\begin{eqnarray}
F & = & F_{\mathrm{mag}}\left[\mathbf{M}_{i}\right]+\alpha_{1}\left(H_{x}^{2}+H_{y}^{2}\right)\left(M_{1}^{2}+M_{2}^{2}\right)\label{Fmag_field}\\
 &  & +\alpha_{2}\left[H_{x}^{2}\left(M_{1,x}^{2}+M_{2,x}^{2}\right)+H_{y}^{2}\left(M_{1,y}^{2}+M_{2,y}^{2}\right)\right]\nonumber \end{eqnarray}
 with coupling constants $\alpha_{1}$ and $\alpha_{2}$. The first
term is isotropic and describes the suppression of SDW order in the
presence of an external field. The second term gives an anisotropic
response. Since $\chi_{\mathrm{mag}}^{jj}\left(\mathbf{q}=0\right)\propto d^{2}F/dH_{j}^{2}$,
we obtain for the anisotropic uniform magnetic susceptibility

\begin{eqnarray}
\chi_{\mathrm{mag}}^{xx}\left(\mathbf{q}=0\right)-\chi_{\mathrm{mag}}^{yy}\left(\mathbf{q}=0\right) & \propto\nonumber \\
\left(M_{1,x}^{2}+M_{2,x}^{2}\right)-\left(M_{1,y}^{2}+M_{2,y}^{2}\right)\label{aux_static_suscept}\end{eqnarray}

To understand the anisotropy in the susceptibility, we cannot use
the $1/N$ (saddle-point) approach discussed in Section \ref{sec_model},
since it does not distinguish between the components of the magnetic
order parameter. Yet, we can gain qualitative understanding by noting
that, in the iron pnictides, the magnetic order parameter points parallel
to the modulation axis in the SDW phase, i.e. $\mathbf{M}_{1}\parallel\hat{x}$
while $\mathbf{M}_{2}\parallel\hat{y}$. Therefore, at least within
a phenomenological approach, there must be a spin-anisotropy term
in the free energy expansion (\ref{F_mag}) favoring these spin directions.
Consequently, the main contribution to the nematic order parameter
$\varphi=M_{1}^{2}-M_{2}^{2}$ comes from $\left\langle M_{1,x}^{2}\right\rangle -\left\langle M_{2,y}^{2}\right\rangle $,
implying that, to leading order in the spin-anisotropy:

\begin{equation}
\chi_{\mathrm{mag}}^{xx}\left(\mathbf{q}=0\right)-\chi_{\mathrm{mag}}^{yy}\left(\mathbf{q}=0\right)\propto\varphi\label{static_suscept}\end{equation}

Interestingly, recent torque magnetometry experiments on $\mathrm{BaFe_{2}As_{2}}$
and $\mathrm{BaFe_{2}\left(As_{1-x}P_{x}\right)_{2}}$ measured the
anisotropy in the uniform magnetic susceptibility, finding a non-zero
value below the structural transition temperature \cite{Matsuda11}.

\subsection{Resistivity anisotropy}

The experimental tool that has been mostly used to probe the nematic
phase is the resistivity anisotropy $\Delta\rho=\rho_{b}-\rho_{a}$
\cite{Chu10,Tanatar10,Kuo11,Blomberg11}. On the theory side, however,
unlike the other properties discussed above, $\Delta\rho$ refers
to a non-equilibrium situation, what brings additional difficulties
to the calculation. Here we follow the same framework of Ref. \cite{Fernandes11}
and employ a semi-classical Boltzmann equation approach to construct
a transport theory for the nematic phase. In this subsection, $b$
and $a$ are parallel to the $y$ and $x$ axis, respectively.

The main property of the nematic phase is that magnetic fluctuations
are different around the two ordering vectors $\mathbf{Q}_{1}=\left(\pi,0\right)$
and $\mathbf{Q}_{2}=\left(0,\pi\right)$. The scattering of electrons
by these anisotropic spin fluctuations will give rise to an anisotropic
scattering rate, resulting in a non-zero resistivity anisotropy $\Delta\rho$
above the magnetic transition temperature $T_{N}$ but below the structural
transition temperature $T_{s}$. Of course, below $T_{N}$ the anisotropic
reconstruction of the Fermi surface by long-range magnetic order will
give another contribution for $\Delta\rho$ \cite{bascones,kotliar},
but here we will focus only on the paramagnetic nematic phase.

The electrons that are most efficiently scattered by spin fluctuations
are the ones near the hot spots of the Fermi surface, $E\left(\mathbf{q}_{hs}\right)=E\left(\mathbf{q}_{hs}+\mathbf{Q}_{i}\right)$,
since they remain near the Fermi level even after acquiring the additional
momentum $\mathbf{Q}_{i}$ transferred in the scattering process.
The electrons on the other parts of the Fermi surface (called cold
regions) do not significantly contribute to the spin-fluctuation scattering.
For a very clean system, it was shown in Ref. \cite{Hlubina95} that
these cold regions of the Fermi surface dominate the transport properties,
short-circuiting the contribution from the hot spots. On the other
hand, Ref. \cite{Rosch99} showed that in dirty systems, the contribution
of the spin-fluctuation scattering is the leading-order correction
to the residual resistivity $\rho_{0}$. Applying these results to
our system, since the resistivity anisotropy comes from the contribution
of the hot spots, we expect $\Delta\rho$ to be larger in samples
that are dirtier. In fact, this has been recently confirmed by measurements
comparing the resistivity anisotropy of normal and annealed samples
\cite{Uchida11}.

To proceed, we consider the dirty limit where $\rho_{0}$ is large
compared to the contribution of the spin-fluctuation scattering. We
also consider a three-band model, with a central circular hole pocket
$\Gamma$ and two elliptical electron pockets $X$ and $Y$ displaced
from the center by the ordering vectors $\mathbf{Q}_{1}=\left(\pi,0\right)$
and $\mathbf{Q}_{2}=\left(0,\pi\right)$, respectively. Due to the
tetragonal symmetry of the system, $X$ transforms to $Y$ under a
$90^{\circ}$ rotation. Within this model, the resistivity along the
$j$ axis is given by (see Ref. \cite{Fernandes11} for more details):

\begin{equation}
\frac{\rho_{jj}-\rho_{0}}{\rho_{0}}=c\sum_{\mathbf{q},\mathbf{q}^{\prime}}\sum_{a=X,Y}\,\frac{\left(v_{\Gamma,\mathbf{q}}^{(j)}-v_{a,\mathbf{q}^{\prime}}^{(j)}\right)^{2}}{\omega_{\mathbf{q}-\mathbf{q}^{\prime}}^{(a)}\left(\omega_{\mathbf{q}-\mathbf{q}^{\prime}}^{(a)}+t\right)}\label{rho_jj}\end{equation}
 where $t\sim T/\gamma$ is a dimensionless temperature, $v_{a,\mathbf{q}}^{(j)}$
is the $j$-component of the Fermi velocity of band $a$, and $c>0$
is a constant proportional to the ratio between the amplitudes of
impurity scattering and spin-fluctuation scattering. Here, $\mathbf{q}$
and $\mathbf{q}^{\prime}$ are momenta restricted to the Fermi surface.
The functions $\omega_{\mathbf{q}-\mathbf{q}^{\prime}}^{(a)}$ can
be written as $\omega_{\mathbf{q}-\mathbf{q}^{\prime}}^{(a)}=\left(r\pm\varphi\right)+\left|\mathbf{q}_{\Gamma}-\mathbf{q}_{a}^{\prime}-\mathbf{Q}_{a}\right|^{2}$,
where the upper (lower) sign refers to band $Y$ ($X$).

\begin{figure}

\begin{centering}
\includegraphics[width=0.9\columnwidth]{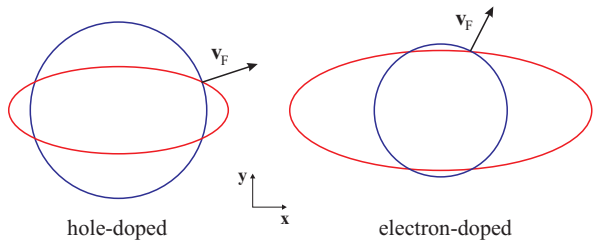} 
\par\end{centering}

\caption{Position of the hot spots in hole-doped and electron-doped compounds.
The elliptical electron pocket (red) was displaced by the ordering
vector $\mathbf{Q}_{1}=\left(\pi,0\right)$, crossing the circular
hole pocket (blue) at the hot spots. The changes in the position of
the hot spots are a consequence of the changes in the areas of the
electron and hole pockets due to the variation in the charge carrier
concentration. We also present the Fermi velocity $\mathbf{v}_{F}=-\mathbf{v}_{\Gamma,\mathbf{q}_{\mathrm{hs}}}$
at the hot spot in each case. Notice that $v_{F}^{(x)}>v_{F}^{(y)}$
in the hole-doped case, whereas $v_{F}^{(y)}>v_{F}^{(x)}$ in the
electron-doped case. According to Eq. (\ref{delta_rho}), one expects
$\Delta\rho<0$ and $\Delta\rho>0$, respectively. }

\label{fig_hot_spots} 
\end{figure}

Performing a $90^{\circ}$ rotation, and using the symmetry properties
relating $\mathbf{v}_{X}$ to $\mathbf{v}_{Y}$, we can expand Eq.
(\ref{rho_jj}) for small $\varphi$ near the structural transition,
obtaining:

\begin{eqnarray}
\frac{\rho_{yy}-\rho_{xx}}{\rho_{0}} & = & c\,\varphi\sum_{\mathbf{q},\mathbf{q}^{\prime}}\left[\left(v_{\Gamma,\mathbf{q}}^{(y)}-v_{X,\mathbf{q}^{\prime}}^{(y)}\right)^{2}-\left(v_{\Gamma,\mathbf{q}}^{(x)}-v_{X,\mathbf{q}^{\prime}}^{(x)}\right)^{2}\right]\nonumber \\
 &  & \times f\left(\mathbf{q}_{\Gamma}-\mathbf{q}_{X}^{\prime}-\mathbf{Q}_{1}\right)\label{delta_rho}\end{eqnarray}
 where the function $f\left(\mathbf{k}\right)$ is given by:

\begin{equation}
f\left(\mathbf{k}\right)=\frac{2\left(2k^{2}+2r+t\right)}{\left(k^{2}+r\right)^{2}\left(k^{2}+r+t\right)^{2}}\label{aux_delta_rho}\end{equation}

Therefore, as expected, $\Delta\rho\sim\varphi$, but the sign of
the proportionality constant can be positive or negative. Also, $\Delta\rho$
may be small even when $\varphi$ is large (or vice-versa) due to
the interplay between impurity-scattering and spin-fluctuation scattering.

The function $f\left(k\right)$ is in general dominated by the $k=0$
contribution - this can be rigorously shown at low temperatures and
near an SDW quantum critical point, where $r/t\rightarrow0$. Thus,
the main contribution to the momentum sum comes from $\mathbf{q},\mathbf{q}^{\prime}$
near $\mathbf{q}_{hs},\mathbf{q}_{hs}+\mathbf{Q}_{1}$, and one can
obtain the sign of the resistivity anisotropy with respect to $\varphi$
by just analyzing the Fermi velocities of the hot spots.

For instance, if the hot spots are close to the $y$ axis, as it is
the case for electron-doped samples \cite{Liu10}, then one expects
the $y$-component of the Fermi velocity to be larger than the $x$-component,
i.e. $\Delta\rho>0$. On the other hand, if the hot spots are close
to the $x$ axis, as it is the case for hole-doped samples, the opposite
holds and one expects $\Delta\rho<0$ (see Fig. \ref{fig_hot_spots}).
Hence, this general model predicts different signs of $\Delta\rho$
(in the paramagnetic phase) for electron-doped and hole-doped systems.
This feature has been recently observed experimentally \cite{Blomberg12},
suggesting that the transport properties in the nematic phase are
well described by this model of anisotropic spin-fluctuation scattering.

We note that there are other contributions that may affect the resistivity
anisotropy in the nematic-paramagnetic phase. For instance, orbital
order (see Section \ref{sec_orbital_order}), which is also induced
by nematic order, changes the Drude weight along the $x$ and $y$
directions. Whether this Drude weight redistribution can account for
the experimentally observed resistivity anisotropy remains an unsettled
issue, as some works find that the resistivity anisotropy coming from
this mechanism alone gives the opposite sign of the $\Delta\rho$
measured experimentally \cite{Phillips10,devereaux10,bascones}. Impurity
effects also play an important role in transport, as discussed above
and also in a different model proposed in Ref. \cite{kontani_nematic}
based on orbital order induced by antiferro-orbital fluctuations.
A recent Monte Carlo simulation in the spin-fermion model also found
that short-range magnetic order (of the same kind that appear in our
nematic model) as well as short-range orbital order can promote anisotropic
transport \cite{Dagotto_Liang}.

\subsection{Elastic modulus: ultra-sound measurements}

So far we discussed several manifestations of long-range nematic order.
An interesting question is how one can also probe nematic fluctuations.
If these are emergent degrees of freedom of the system, then we expect
that their low-energy excitations will have an important effect in
the normal state properties.

As we discussed in Section \ref{sec_model}, nematic fluctuations
are four-spin correlation functions, which are rather difficult to
measure using standard magnetic probes. An alternative is to investigate
how they couple to other degrees of freedom. For instance, Eq. (\ref{F_el})
shows that the nematic order parameter acts as a conjugate field to
the orthorhombic distortion. Although in the tetragonal phase the
mean value of the nematic order parameter is zero, nematic fluctuations
are still present. As a result, they will give rise to a quadratic
term in the shear distortion $\varepsilon_{s}$ that effectively reduces
the shear modulus $C_{s}$. Physically, nematic fluctuations suppress
the energy cost of a momentaneous orthorhombic distortion, what is
translated as a softening of $C_{s}$. Quantitatively, the renormalized
shear modulus is given by:

\begin{equation}
\tilde{C}_{s}^{-1}=\lim_{\sigma\rightarrow0}\left(\frac{\partial^{2}\ln Z}{\partial\sigma^{2}}\right)\label{definition_Cs}\end{equation}
 where $Z=\int dM_{i}\, d\varepsilon_{s}\mathrm{e}^{-F_{\mathrm{mag}}-F_{\mathrm{el}}}$
is the partition function, see Eqs. (\ref{F_mag}) and (\ref{F_el_stress}).
For a harmonic lattice, the elastic degrees of freedom can be integrated
out from the partition function analytically, and we obtain $Z=\int dM_{i}\,\mathrm{e}^{-\tilde{F}_{\mathrm{mag}}}$,
with $\tilde{F}_{\mathrm{mag}}$ given by Eq. (\ref{F_mag_el_renormalized}).
Taking the derivatives with respect to the infinitesimal stress $\sigma$
and then making $\sigma\rightarrow0$ we obtain \cite{FernandesPRL10}:

\begin{eqnarray}
\tilde{C}_{s}^{-1} & = & C_{s}^{-1}+\frac{\lambda_{\mathrm{el}}^{2}}{C_{s}^{2}}\,\chi_{\mathrm{nem}}\nonumber \\
\left(\frac{\tilde{C}_{s}}{C_{s}}\right)^{-1} & = & 1+\frac{\lambda_{\mathrm{el}}^{2}}{C_{s}}\,\chi_{\mathrm{nem}}\label{Cs_renormalized}\end{eqnarray}
 where $\chi_{\mathrm{nem}}\left(q\right)=\left\langle \left(M_{1}^{2}-M_{2}^{2}\right)_{q}\left(M_{1}^{2}-M_{2}^{2}\right)_{-q}\right\rangle $,
as defined in Section \ref{sec_model}. Recall that:

\begin{equation}
\chi_{\mathrm{nem}}=\frac{\int_{q}\chi^{2}\left(q\right)}{1-\tilde{g}\int_{q}\chi^{2}\left(q\right)}\label{repeat_nematic_suscept}\end{equation}
 with renormalized $\tilde{g}=g+\lambda_{\mathrm{el}}^{2}/C_{s}$
(see Eq. \ref{F_mag_el_renormalized}). As expected, the divergence
of the nematic susceptibility leads to the vanishing of the shear
modulus.

Most interestingly, Eq. (\ref{Cs_renormalized}) shows that one can
experimentally extract $\chi_{\mathrm{nem}}$ by just measuring the
shear modulus $\tilde{C}_{s}$ via standard ultra-sound techniques
\cite{FernandesPRL10,Yoshizawa12}. To calculate $\chi_{\mathrm{nem}}$,
one can use inelastic neutron scattering data as input, since they
provide the parameters of $\chi\left(q\right)=\xi^{-2}+q^{2}+\gamma^{-1}\left|\omega_{n}\right|$.
With this procedure, it is possible to establish whether the measured
softening of $\tilde{C}_{s}$ agrees with what one would expect if
the structural transition was driven by magnetic fluctuations. This
procedure was carried on in Ref. \cite{FernandesPRL10}, and a very
good agreement was found between the measured $\tilde{C}_{s}$ and
the calculated $\chi_{\mathrm{nem}}$, based on neutron scattering
data.

\begin{figure}

\begin{centering}
\includegraphics[width=0.9\columnwidth]{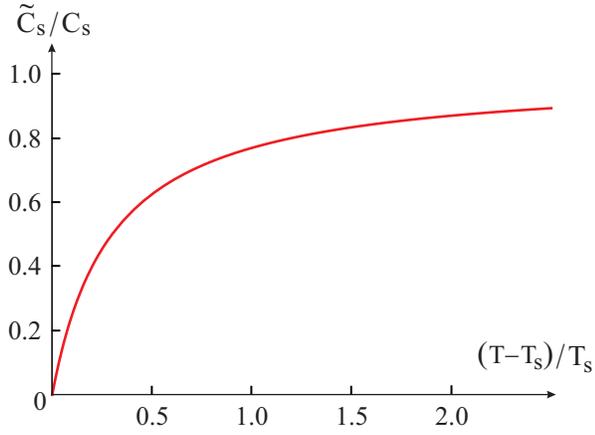} 
\par\end{centering}

\caption{Typical behavior of the shear modulus in the tetragonal phase as function
of the reduced temperature. The softening follows Eq. (\protect\ref{softening_shear_modulus});
here we used $\kappa=0.3$. The quantity $\tilde{C_{s}}$ is the physical
shear modulus, that can be measured in ultra-sound experiments, while
$C_{s}$ is its high-temperature limit. }

\label{fig_shear_modulus} 
\end{figure}

For a finite-temperature nematic transition, $\int_{q}\chi^{2}\left(q\right)\sim\xi^{4-d}\sim\left(T-T_{N,0}\right)^{-\left(4-d\right)\nu}$.
Setting $d=2$ and using the mean-field exponent $\nu=1/2$, we obtain
a simple expression for the vanishing of the shear modulus:

\begin{equation}
\tilde{C}_{s}=C_{s}\left(1+\kappa\frac{T_{s}}{T-T_{s}}\right)^{-1}\label{softening_shear_modulus}\end{equation}
 where $\kappa$ is a positive dimensionless constant, of the order
of $\kappa\sim\frac{\lambda_{\mathrm{el}}^{2}}{C_{s}\left(E_{\mathrm{mag}}+\tilde{g}\right)}$,
with $E_{\mathrm{mag}}$ denoting a characteristic magnetic energy
scale. Notice that the shear modulus is reduced to half of its saturation
value at $T=T_{s}\left(1+\kappa\right)$. The typical behavior of
this function is shown in Fig. \ref{fig_shear_modulus} and agrees
qualitatively well with the experimentally observed softening of $\tilde{C}_{s}$
\cite{FernandesPRL10,Yoshizawa12}.

\section{Nematic order or orbital order? \label{sec_orbital_order}}

Besides nematic order, it has been proposed that ferro-orbital order
could drive the tetragonal symmetry-breaking observed in the iron
pnictides \cite{kruger09,RRPSingh09,w_ku10,Phillips10,devereaux10,Nevidomskyy11}.
In this case, the occupation of the Fe $d_{xz}$ orbital is different
than the occupation of the $d_{yz}$ orbital. Since the total occupation
number depends on all states below the Fermi level, in some cases
it is more convenient to refer to the splitting between the on-site
orbital energies, $\Delta_{\mathrm{orb}}\equiv\Delta_{xz}-\Delta_{yz}\neq0$.
ARPES measurements have observed $\Delta_{\mathrm{orb}}<0$ in detwinned
electron-doped $\mathrm{Ba\left(Fe_{1-x}Co_{x}\right)_{2}As_{2}}$
at the same temperature where a non-zero resistivity anisotropy sets
in.

By symmetry, the ferro-orbital order parameter enters the free energy
via:

\begin{equation}
F_{\mathrm{orb}}=\int_{x}\frac{\chi_{\mathrm{orb}}^{-1}}{2}\,\Delta_{\mathrm{orb}}^{2}-\lambda_{\mathrm{orb}}\int_{x}\Delta_{\mathrm{orb}}\left(M_{1}^{2}-M_{2}^{2}\right)\label{F_orb}\end{equation}
 where\textbf{ $\chi_{\mathrm{orb}}\equiv\chi_{\mathrm{orb}}\left(q=0\right)$
}is the intrinsic ferro-orbital susceptibility, with\textbf{ $\chi_{\mathrm{orb}}\left(q\right)=\left\langle \Delta_{\mathrm{orb}}\left(q\right)\Delta_{\mathrm{orb}}\left(-q\right)\right\rangle $},
and $\lambda_{\mathrm{orb}}$ is the appropriate coupling constant
to the magnetic degrees of freedom. By minimizing the free energy,
it follows that $\Delta_{\mathrm{orb}}\propto\lambda_{\mathrm{orb}}\varphi$.
In Ref. \cite{FernandesPRB12}, the coupling constant $\lambda_{\mathrm{orb}}$
was calculated using an itinerant approach, and was shown to be negative
for electron-doped compounds, providing a possible explanation for
why the ARPES measurements observed $\Delta_{\mathrm{orb}}<0$ when
$\varphi>0$.

Thus, within the nematic scenario, orbital order is induced by long-range
nematic order. However, one could argue that in a model with spontaneous
orbital order, where $\chi_{\mathrm{orb}}^{-1}\propto T-T_{\mathrm{orb}}$,
orbital order is the one that drives the nematic order. Of course,
a phenomenological approach based only on symmetry considerations
cannot distinguish between these two different scenarios. One then
has to rely on model microscopic calculations to investigate when
spontaneous orbital order and/or spontaneous nematic order appear.

In what concerns the nematic scenario, as discussed in Section \ref{sec_model},
a spontaneous tendency to long-range nematic order is found in both
strong-coupling (the $J_{1}-J_{2}$ localized-spin model) and weak-coupling
limits (the multi-band nested electronic model). On the other hand,
spontaneous ferro-orbital order is mostly found in particular (but
not all, see Ref. \cite{kruger09}) versions of the Kugel-Khomskii
model, which is obtained in the strong-coupling limit. Furthermore,
different numerical analyses of the five-band Hubbard model did not
find a ground state that displays orbital order in the absence of
long-range magnetic order \cite{bascones,dagotto,kotliar}. Although
they do not constitute a proof, these results, combined with the fact
that the magnetism in the iron pnictides is closer to the itinerant
rather than to the localized-spin regime (since these systems are
metallic, see also optical conductivity measurements \cite{Wang08}),
suggest that the nematic instability may be the one driving orbital
order and structural order.

This does not imply that the role played by the orbital degrees of
freedom is irrelevant: for example, it is clear that the reconstruction
of the Fermi surface due to orbital order improves the nesting conditions
and enhances the magnetic transition temperature \cite{Phillips10,ZXShen_NaFeAS}.
The orbital character of the different pockets of the Fermi surface
is also important for the superconducting state \cite{Zhang09}. Nevertheless,
orbital order alone cannot explain the observed anisotropic properties
of the iron pnictides. For instance, calculations of the resistivity
anisotropy based on the experimentally observed $\Delta_{\mathrm{orb}}<0$
give the opposite sign of the experimental $\rho_{b}-\rho_{a}$ \cite{Phillips10,devereaux10,bascones},
indicating that the scattering by anisotropic spin fluctuations is
fundamental to explain the observed resistivity anisotropy.

Rather than competing tendencies, ferro-orbital and nematic order
are actually cooperative instabilities, which can potentially lead
to the enhancement of both the structural transition temperature and
the anisotropic properties of the orthorhombic state. To our knowledge,
a model that displays both spontaneous nematic \emph{and} orbital
order has not been conceived so far.

Nevertheless, it is clear that there is a coupling between the orbital
and magnetic degrees of freedom, since first-principle calculations
find that the onset of long-range SDW order is accompanied by ferro-orbital
order, due to the different spin polarizations of the $d_{xz}$ and
$d_{yz}$ orbitals \cite{kotliar}. If we assume that the system has
a tendency to orbital order, but that $\chi_{\mathrm{orb}}^{-1}>0$
at all temperatures, then we can integrate out the orbital degrees
of freedom from the partition function, obtaining the following expression
for the nematic susceptibility (see Eq. \ref{repeat_nematic_suscept}):

\begin{equation}
\chi_{\mathrm{nem}}^{-1}=\left[\int_{q}\chi_{\mathrm{mag}}^{2}\left(q\right)\right]^{-1}-g-\lambda_{\mathrm{orb}}^{2}\,\chi_{\mathrm{orb}}-\lambda_{\mathrm{el}}^{2}\, C_{s}^{-1}\label{chi_nem_orb}\end{equation}

It is clear from this expression how the different degrees of freedom
assist nematic fluctuations and nematic order. The first term is just
the usual Gaussian fluctuations of the magnetic order parameter, going
to zero only at the (bare) magnetic transition temperature. The second
term $g>0$ also comes from the magnetic sector (see Eq. \ref{F_mag})
and is responsible for the degeneracy of the two magnetic stripe states,
i.e. it is the {}``intrinsic'' magnetic instability towards a nematic
state. The third term is associated with the orbital degrees of freedom
and shows that a tendency towards ferro-orbital order enhances the
nematic susceptibility. Finally, the last term, which originates from
the magneto-elastic coupling, implies that softer lattices have a
larger effective nematic coupling.

\section{Interplay between nematicity and superconductivity \label{sec_nematic_SC}}

In the previous sections, we explored the effects of the nematic degrees
of freedom on the normal state properties. Here, we investigate how
superconductivity and nematic order/fluctuations affect each other.

\subsection{Competition with superconducting order}

It has been established both experimentally and theoretically that
SC and SDW are competing orders in the iron pnictides. Neutron diffraction
measurements observe a strong suppression of the SDW order parameter
below $T_{c}$ - in some materials, there is even a reentrance of
the non-magnetic phase at low temperatures \cite{FernandesPRB10}.
Theoretically, the fact that the same electrons are responsible for
both the static staggered magnetization and the superfluid density
explains such a strong competition \cite{Fernandes_Schmalian,Vorontsov10}.
In our phenomenological approach, we can capture this competition
by adding to the free energy expansion a bi-quadratic SDW-SC term:

\begin{equation}
F\left[\mathbf{M}_{i},\Delta\right]=F_{\mathrm{mag}}\left[\mathbf{M}_{i}\right]+F_{\mathrm{SC}}\left[\Delta\right]+\lambda_{\mathrm{sc}}\left(M_{1}^{2}+M_{2}^{2}\right)\Delta^{2}\label{F_SC}\end{equation}

In the mean-field level, a non-zero SC order parameter renormalizes
the magnetic susceptibility, suppressing the SDW transition temperature
$T_{N,0}$:

\begin{equation}
r_{0}\rightarrow r_{0}+\lambda_{\mathrm{sc}}\Delta^{2}\label{r0_SC}\end{equation}

Within this phenomenological approach, it is straightforward to understand
qualitatively the effects of superconducting order on the nematic
order parameter. Below $T_{c}$, the magnetic susceptibility is suppressed
according to Eq. (\ref{r0_SC}). Since long-range nematic order is
induced by magnetic fluctuations, it will also be suppressed by SC.
This implies that both states are also competing orders, despite the
fact that there is no direct coupling between the nematic and SC order
parameters.

\begin{figure}

\begin{centering}
\includegraphics[width=0.95\columnwidth]{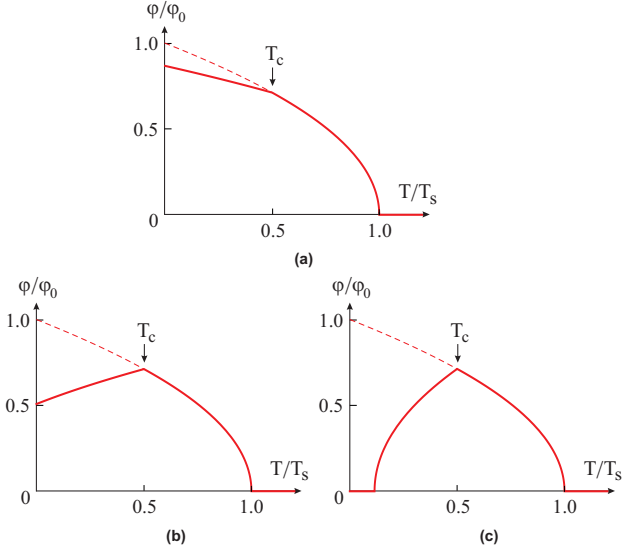} 
\par\end{centering}

\caption{Nematic order parameter $\varphi$ (in units of its zero-temperature
value $\varphi_{0}$) as a function of temperature (in units of the
structural transition temperature $T_{s}$). The dashed line corresponds
to the case with no SC, while the solid line corresponds to the solution
in the presence of a SC state that sets in at $T_{c}=T_{s}/2$. The
SDW-SC coupling constant was set to $\frac{\tilde{\lambda}_{\mathrm{sc}}\Delta_{0}^{2}}{\alpha}=0.5$
(a), $1.5$ (b) and $2.3$ (c). }

\label{fig_nematic_SC} 
\end{figure}

To investigate the interplay between nematicity and SC quantitatively,
we consider the case where the system has no long-range magnetic order
once SC sets in. First we analyze the behavior of the nematic order
parameter $\varphi$, i.e. we assume $T_{c}<T_{s}$. For simplicity,
we solve the self-consistent equations (\ref{self_cons_eq}) in $d=2$,
since in this case there is no finite-temperature SDW transition.
A straightforward calculation leads to an implicit equation for $\tilde{\varphi}=4\pi\varphi/gT_{N}$
(see Ref. \cite{FernandesPRB12}):

\begin{equation}
\tilde{\varphi}\coth\tilde{\varphi}+\frac{u}{g}\,\ln\left(\frac{\tilde{\varphi}}{\sinh\tilde{\varphi}}\right)=\tilde{r}_{0}+\tilde{\lambda}_{\mathrm{sc}}\Delta^{2}\label{phi_Delta}\end{equation}
 where $\tilde{r}_{0}=\alpha\left(T-T_{N}\right)$ and $\tilde{\lambda}_{\mathrm{sc}}=4\pi\lambda_{\mathrm{sc}}/gT_{N}$.
For a fixed value of the ratio $u/g$, it is straightforward to solve
the self-consistent equation (\ref{phi_Delta}) for $\tilde{\varphi}$.
In mean-field, for $T>T_{c}$, we have $\Delta^{2}=0$, whereas for
$T<T_{c}$, it holds that $\Delta^{2}=\Delta_{0}^{2}\left(T_{c}-T\right)$.
The structural transition temperature takes place at $T_{s}=T_{N}+\alpha^{-1}$.
Using Eq. (\ref{phi_Delta}), we can evaluate $d\varphi/dT$ for all
temperatures, obtaining:

\begin{eqnarray}
\left(\frac{d\varphi}{dT}\right)_{T_{c}<T<T_{s}} & = & -c\nonumber \\
\left(\frac{d\varphi}{dT}\right)_{T<T_{c}\phantom{<T_{s}}} & = & -\left(1-\frac{\tilde{\lambda}_{\mathrm{sc}}\Delta_{0}^{2}}{\alpha}\right)c\label{derivative}\end{eqnarray}
with $c>0$ a temperature-dependent positive number. Eqs. (\ref{derivative})
show that the nematic order parameter is suppressed below $T_{c}$.
If the competition between SDW and SC is strong enough, such that
$\tilde{\lambda}_{\mathrm{sc}}\Delta_{0}^{2}>\alpha$, then $\frac{d\varphi}{dT}$
changes sign below $T_{c}$, implying that the orthorhombic distortion
actually decreases inside the SC state. This opens up the possibility
of completely killing the Ort phase at low temperatures, promoting
a reentrance of the Tet phase. In Fig. \ref{fig_nematic_SC}, we illustrate
these behaviors by plotting the solutions of Eq. (\ref{phi_Delta})
as a function of temperature for different values of the coupling
constant $\tilde{\lambda}_{\mathrm{sc}}$. The competition between
the two order parameters has also been investigated by Ref. \cite{Moon11},
where a different approach was employed.

Remarkably, x-ray diffraction measurements have found a very strong
suppression of the orthorhombic distortion inside the SC phase, and
even a reentrance of the tetragonal phase for $\mathrm{Ba\left(Fe_{1-x}Co_{x}\right)_{2}As_{2}}$
near optimal doping \cite{Nandi09}. This behavior was observed in
the absence of long-range magnetic order, when only the SC and nematic
order parameters are non-zero. These measurements provide yet another
strong evidence for the electronic character of the structural transition
in the iron pnictides, and find a natural explanation within the nematic
scenario. No additional coupling between the nematic and SC order
parameters is necessary, as the magnetic origin of the nematic phase,
combined with the competition between SDW and SC, give rise to an
indirect competition between nematicity and SC.

This competition is also manifested in the shear modulus $\tilde{C}_{s}$.
An interesting case to analyze is when the system does not have long-range
nematic or magnetic order, but only SC at a finite $T_{c}$ (as it
is the case in optimally doped samples). For concreteness, we consider
the presence of a SDW quantum critical point, such that $r\propto T$.
To calculate the shear modulus, we need to evaluate $\int_{q}\chi_{\mathrm{mag}}^{2}\left(q\right)$
in Eq. (\ref{repeat_nematic_suscept}). Above $T_{c}$, we have the
usual overdamped spin excitations, $\chi_{\mathrm{mag}}^{-1}\left(q\right)=r+q^{2}+\gamma^{-1}\left|\omega_{n}\right|$.
Considering for simplicity $d=2$ and the $T=0$ limit, we obtain:

\begin{equation}
\chi_{\mathrm{nem}}^{-1}=\left[\frac{\gamma}{4\pi^{2}}\ln\left(\frac{\Lambda}{r\gamma}\right)\right]^{-1}-\tilde{g}\label{chi_nem_aboveTc}\end{equation}
where $\Lambda$ is the frequency cutoff. Below $T_{c}$, there are
changes not only in the static part of the susceptibility, $r\rightarrow\tilde{r}=r+\lambda_{\mathrm{sc}}\Delta^{2}$,
but also in the dynamics of $\chi_{\mathrm{mag}}\left(q\right)$.
The presence of a SC gap makes the spin dynamics ballistic for energies
$\omega_{n}\ll\Delta$, while for $\omega_{n}\gg\Delta$ one recovers
the overdamped dynamics \cite{Abanov99}. To capture these two different
behaviors, we write the phenomenological form $\chi_{\mathrm{mag}}\left(q\right)=r+q^{2}+f\left(\omega_{n}\right)$
with $f\left(\omega_{n}\right)=\omega_{n}^{2}/\left(\Delta\gamma\right)$
for $\omega_{n}<\Delta$ and $f\left(\omega_{n}\right)=\omega_{n}/\gamma$
for $\omega_{n}>\Delta$. A straightforward calculation then gives:

\begin{figure}

\begin{centering}
\includegraphics[width=0.9\columnwidth]{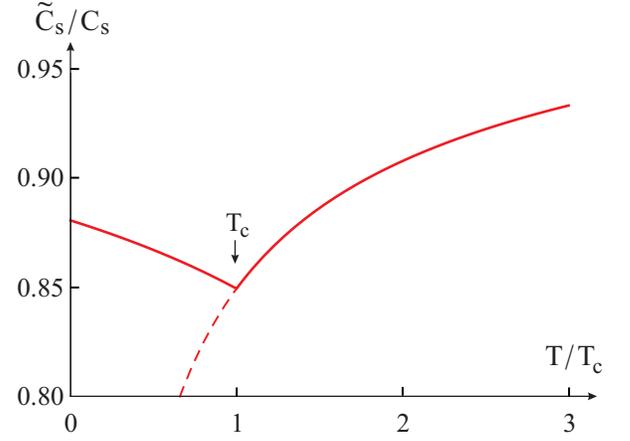} 
\par\end{centering}

\caption{Shear modulus $\tilde{C}_{s}$ (in units of its high-temperature value
$C_{s}$) as a function of temperature (in units of $T_{c}$). The
dashed line refers to the case without SC and with a putative SDW
quantum critical point. The solid line corresponds to the case where
SC sets in at $T_{c}$, and reveals the hardening of the shear modulus
inside the Tet-SC phase. }

\label{fig_shear_modulus_SC} 
\end{figure}

\begin{equation}
\chi_{\mathrm{nem}}^{-1}=\frac{4\pi^{2}}{\gamma}\left[\sqrt{\frac{\Delta}{\tilde{r}\gamma}}\arctan\left(\sqrt{\frac{\Delta}{\tilde{r}\gamma}}\right)+\ln\left(\frac{\Lambda}{\tilde{r}\gamma+\Delta}\right)\right]^{-1}-\tilde{g}\label{chi_nem_belowTc}\end{equation}
 with $\tilde{r}=r+\lambda_{\mathrm{sc}}\Delta^{2}$. Substituting
this expression in Eq. (\ref{Cs_renormalized}), we obtain the renormalized
shear modulus. In Fig. \ref{fig_shear_modulus_SC}, we plot $\tilde{C}_{s}$
using parameters such that $\gamma\gg\Delta$, in accordance to what
neutron scattering experiments observe. Notice that the shear modulus
is enhanced below $T_{c}$, i.e. the lattice is hardened in the SC
phase, due to the suppression of nematic fluctuations. We emphasize
that the origin of this behavior is in the magnetic nature of the
nematic degrees of freedom, allied to the competition between SDW
and SC. The hardening of the shear modulus below $T_{c}$ has been
observed by ultra-sound measurements in optimally doped $\mathrm{Ba\left(Fe_{1-x}Co_{x}\right)_{2}As_{2}}$
\cite{FernandesPRL10,Yoshizawa12}, providing another evidence in
favor of the nematic model.

\subsection{Nematic fluctuations and superconductivity}

Given the evidence in favor of nematic degrees of freedom in the iron
pnictides, it is natural to ask whether nematic fluctuations are able
to enhance the value of $T_{c}$. Here we give a qualitative argument
that this may be the case in the iron pnictides, although more rigorous
calculations are necessary to confirm it. We start with a simplified
model containing one hole pocket and one electron pocket, both with
the same density of states $N_{F}$. The linearized BCS equations
for this two-band problem is given by the matrix equation \cite{maiti}:

\begin{equation}
\left(\begin{array}{c}
\Delta_{1}\\
\Delta_{2}\end{array}\right)=N_{F}\ln\left(\frac{W}{T_{c}}\right)\left(\begin{array}{cc}
-U_{1} & -V\\
-V & -U_{2}\end{array}\right)\left(\begin{array}{c}
\Delta_{1}\\
\Delta_{2}\end{array}\right)\label{BCS_two_bands}\end{equation}
 where $W$ is the typical energy scale of the pairing interaction,
$U_{1}$ and $U_{2}$ are intra-band interactions, and $V$ is the
inter-band interaction. The value of $T_{c}$ is given by the largest
eigenvalue $\lambda_{+}$ of the matrix:

\begin{equation}
T_{c}=W\,\mathrm{e}^{-1/\left(N_{F}\lambda_{+}\right)}\label{Tc}\end{equation}

If $U_{1}$ and $U_{2}$ are repulsive interactions, i.e. $U_{i}>0$,
then SC will only appear if $\left|V\right|>\sqrt{U_{1}U_{2}}$. The
largest eigenvalue is given by:

\begin{equation}
\lambda_{+}=-\left(\frac{U_{1}+U_{2}}{2}\right)+\sqrt{\left(\frac{U_{1}-U_{2}}{2}\right)^{2}+V^{2}}\label{eigenvalue}\end{equation}
 and the eigenvectors give $\Delta_{1}$ and $\Delta_{2}$ with the
same sign if $V<0$ (attractive) or $\Delta_{1}$ and $\Delta_{2}$
with opposite signs if $V>0$ (repulsive). In the first case, the
SC state is called $s^{++}$ state, while in the second case it is
called $s^{+-}$ state. In either case, the interband interaction
must be large enough to overcome the intraband repulsion. In the $s^{+-}$
state this becomes possible due to the proximity to a SDW instability,
which enhances the interband repulsion $V$ \cite{magnetic}. The
$s^{++}$ state is the leading instability if interband orbital fluctuations
(i.e. anti-ferro orbital fluctuations) are strong enough to enhance
the electron-phonon attractive interaction \cite{orbital}. Several
experimental results indicate that the state realized in the iron
pnictides is the $s^{+-}$. Evidence includes the existence of a resonance
mode in the magnetic excitation spectrum below $T_{c}$ \cite{Christianson08};
the pattern of quasi-particle interference in the presence of a magnetic
field \cite{Hanaguri10}; half flux-quantum transitions in composite
loops of Nb-iron pnictides \cite{flux_jumps}; and the microscopic
coexistence between SC and SDW \cite{FernandesPRB10}.

How nematic fluctuations affect this model? The key point is that,
unlike the SDW fluctuations, nematic fluctuations are peaked at $q=0$.
Therefore, their main effect should be on the intraband component
of the interaction.\textbf{ }Moreover, unlike magnetic fluctuations,
nematic fluctuations couple to the charge of the electron\textbf{s.
}As such, their coupling the electrons is analogous to the coupling
between the latter and acoustic phonons, resulting in an attractive
intraband interaction $U_{\mathrm{nem}}<0$, i.e. $U_{1}\rightarrow U_{1}-\left\vert U_{\mathrm{nem}}\right\vert $.
Then, according to Eq. (\ref{eigenvalue}), the largest eigenvalue
$\lambda_{+}$ would increase and, consequently, $T_{c}$ would be
enhanced.\textbf{ }Notice that since the pairing interaction due to
nematic fluctuations has an intra-band character, it affects $T_{c}$
in the same way regardless of whether the ground state is $s^{++}$
or $s^{+-}$.

Of course, further calculations are necessary to confirm this qualitative
picture. One argument that supports a rather weak additional role
of nematic-pairing is the absence of significant variations in the
gap amplitude on the hole pocket\textbf{ }(although recent measurements
in LiFeAs found angular variations in the hole-pockets gaps, see \cite{LiFeAs_gap}).
Notice that the nematic fluctuations would always couple to the form
factor $\left(k_{x}^{2}-k_{y}^{2}\right)^{2}$, which should give
rise to smaller pairing gaps on the Brillouin zone diagonals. Also,
key points neglected in this analysis are the possibility of different
frequency cutoffs $W$ among the different interactions and the fact
that the nematic coupling should be treated as a quantum critical
pairing where all energy scales of the problem contribute to the pairing
interaction, not only the ones that are below the typical excitation
energy of the nematic mode \cite{Abanov01}. Yet, since nematic fluctuations
are generated by magnetic fluctuations, it is certainly an interesting
open problem to investigate whether this novel collective mode can
play a favorable role in Cooper pairing.

\section{Concluding remarks \label{sec_conclusion}}

In this paper we discussed distinctive manifestations of the nematic
degrees of freedom in several properties of the iron pnictides. Due
to the magnetic origin of the nematic phase, magnetic properties such
as the spin-lattice relaxation rate, the spin-spin correlation function,
and the uniform magnetic susceptibility display characteristic features
when the nematic order parameter becomes non-zero at $T_{s}$. Furthermore,
the enhancement of magnetic fluctuations at $T_{s}$ also has consequences
to the electronic spectrum, opening a pseudogap in the electronic
spectral function. Scattering of electrons by these anisotropic spin
fluctuations in the nematic paramagnetic phase leads to a resistivity
anisotropy whose sign is different in electron-doped and hole-doped
compounds.

We also showed that the magneto-elastic coupling makes the orthorhombic
distortion proportional to the nematic order parameter, and the shear
modulus proportional to the inverse nematic susceptibility. In the
normal state, the shear modulus is softened by low-energy nematic
fluctuations, but in the SC state it becomes harder due to the indirect
competition between nematicity and SC. This indirect competition,
rooted on the direct competition between SDW and SC for the same electronic
states, is also manifested in the suppression of the orthorhombic
distortion below $T_{c}$. The role of orbital order to the onset
of long-range nematic order was also discussed. We showed that even
when orbital order is not a spontaneous instability of the system,
ferro-orbital fluctuations enhance the nematic susceptibility, whose
divergence then gives rise to a finite orbital polarization.

It is important to emphasize that, besides orbital order and nematic
order, other mechanisms have been proposed for the electronic tetragonal
symmetry-breaking in the pnictides. In Ref. \cite{zlatko}, the authors
argue that an orbital-current density wave may be spontaneously generated
due to the nesting features of the Fermi surface. As a secondary effect
of this orbital-current state, a charge-density wave appears, triggering
the structural transition. The authors of Refs. \cite{Vojta10,Laad11},
on the other hand, suggest proximity to an orbital-selective Mott
transition as the origin of the orthorhombic state. Kontani \emph{et
al.} propose an electronic nematic state rooted on quadrupolar orbital
interactions allied to impurity effects \cite{kontani_nematic}. In
Ref. \cite{daghofer_nematic}, Daghofer \emph{et al}. performed a
numerical study of a three-band Hubbard model and found nematic features
in the electronic spectral function, although the tetragonal symmetry
of the model was explicitly broken by an anisotropic exchange constant.
In Ref. \cite{Indranil11}, Paul proposes a mechanism for the structural
transition based on the magneto-elastic coupling, which is very similar
to the emergent nematic model discussed here. We note also that some
works treat the coupling between the SDW and structural transitions
phenomenologically \cite{Cano10}. Such an approach has the clear
advantage of providing model-independent predictions, but it does
not address why these two transitions follow each other in the phase
diagram.

Several of the manifestations of the nematic phase discussed here
have been detected experimentally - such as the suppression of the
orthorhombic distortion below $T_{c}$ (x-ray diffraction) \cite{Nandi09};
the enhancement of $1/T_{1}T$ below $T_{s}$ (NMR) \cite{Ma_NaFeAs};
the anisotropy in the uniform susceptibility (torque magnetometry)
\cite{Matsuda11}; the softening of the shear modulus in the normal-Tet
phase and its hardening in the SC-Tet phase (ultrasound measurements)
\cite{FernandesPRL10,Yoshizawa12}; and the change in the sign of
the resistivity anisotropy upon hole-doping (transport)\cite{Blomberg12}.
Yet, there are two important predictions of the nematic model that
remain to be seen: the inequivalence between magnetic fluctuations
around the ordering vectors $\mathbf{Q}_{1}=\left(\pi,0\right)$ and
$\mathbf{Q}_{2}=\left(0,\pi\right)$ below $T_{s}$ (inelastic neutron
scattering) and the possible opening of a pseudogap in the electronic
spectrum below $T_{s}$ (ARPES and STM). Other manifestations of nematic
order not presented here have also been discussed in the literature,
such as the effects of nematicity on the SC vortex structure \cite{Song11,Sachdev_nematic11},
on the elastic domain walls \cite{cano_domains}, and on the scattering
by impurities \cite{Davis10}.

Overall, the nematic model offers a simple framework to understand
the interplay between the several degrees of freedom present in the
phase diagrams of the iron pnictides, shedding light on the primary
role played by magnetism. An interesting topic that deserves further
attention is the importance of nematic fluctuations to high-temperature
SC. Here, we gave a qualitative argument of how $T_{c}$ can be enhanced
by nematic fluctuations. A detailed investigation of this result can
potentially have implications also for other systems where nematicity
has been proposed, such as some cuprates \cite{kivelson_cuprates}
and some heavy fermion systems \cite{matsuda_hidden_order}.

We would like to thank E. Abrahams, M. Allan, J. Analytis, E. Bascones,
A. Böhmer, B. Büchner, P. Canfield, P. Chandra, A. Chubukov, J.-H.
Chu, L. Degiorgi, I. Eremin, I. Fisher, A. Goldman, V. Keppens, J.
Knolle, A. Kreyssig, R. McQueeney, D. Mandrus, Y. Matsuda, I. Mazin,
C. Meingast, A. Millis, S. Nandi, I. Paul, D. Pratt, R. Prozorov,
T. Shibauchi, M. Tanatar, and Z. Tesanovic for fruitful discussions.
R.F.M. is supported by the NSF Partnerships for International Research
and Education (PIRE) program OISE-0968226.


\begin{thebibliography}{10}
\bibitem{reviews} D. C. Johnston, Adv. Phys. \textbf{59}, 803 (2010);
J. Paglione and R. L. Greene, Nature Phys. \textbf{6}, 645 (2010).
P. J. Hirschfeld, M. M. Korshunov, and I. I. Mazin, Rep. Prog. Phys.
\textbf{74}, 124508 (2011); D. N. Basov and A. V. Chubukov, Nature
Phys. \textbf{7}, 241 (2011); P. C. Canfield and S. L. Bud'ko, Annu.
Rev. Cond. Mat. Phys. \textbf{1}, 27 (2010); H. H. Wen and S. Li,
Annu. Rev. Cond. Mat. Phys. \textbf{2}, 121 (2011); A. V. Chubukov,
Annu. Rev. Cond. Mat. Phys. \textbf{3}, 57 (2012); G. R. Stewart,
Rev. Mod. Phys. \textbf{83} 1589 (2011).

\bibitem{magnetic} I. I. Mazin, D. J. Singh, M. D. Johannes, and
M. H. Du, Phys. Rev. Lett. \textbf{101}, 057003 (2008); K. Kuroki,
S. Onari, R. Arita, H. Usui, Y. Tanaka, H. Kontani, and H. Aoki, Phys.
Rev. Lett. \textbf{101}, 087004 (2008); A. V. Chubukov, D. V. Efremov,
and I. Eremin. Phys. Rev. B, \textbf{78}, 134512 (2008); V. Cvetkovic
and Z. Tesanovic, EPL \textbf{85}, 37002 (2009); R. Sknepnek, G. Samolyuk,
Y. Lee, and J. Schmalian, Phys. Rev. B \textbf{79}, 054511 (2009);
A.F. Kemper, T.A. Maier, S. Graser, H-P. Cheng, P.J. Hirschfeld and
D.J. Scalapino, New J. Phys. \textbf{12}, 073030 (2010).

\bibitem{FernandesPRB10} R. M. Fernandes, D. K. Pratt, W. Tian, J.
Zarestky, A. Kreyssig, S. Nandi, M. G. Kim, A. Thaler, N. Ni, P. C.
Canfield, R. J. McQueeney, J. Schmalian, and A. I. Goldman, Phys.
Rev. B \textbf{81}, 140501(R) (2010).

\bibitem{Nandi09} S. Nandi, M. G. Kim, A. Kreyssig, R. M. Fernandes,
D. K. Pratt, A. Thaler, N. Ni, S. L. Bud'ko, P. C. Canfield, J. Schmalian,
R. J. McQueeney, and A. I. Goldman, Phys. Rev. Lett. \textbf{104},
057006 (2010).

\bibitem{Fisher11} I. R. Fisher, L. Degiorgi, and Z. X. Shen, Rep.
Prog. Phys. \textbf{74} 124506 (2011).

\bibitem{Chu10} J.-H. Chu, J. G. Analytis, K. De Greve, P. L. McMahon,
Z. Islam, Y. Yamamoto, and I. R. Fisher, Science \textbf{329}, 824
(2010).

\bibitem{Tanatar10} M. A. Tanatar, E. C. Blomberg, A. Kreyssig, M.
G. Kim, N. Ni, A. Thaler, S. L. Bud'ko, P. C. Canfield, A. I. Goldman,
I. I. Mazin, and R. Prozorov, Phys. Rev. B \textbf{81}, 184508 (2010).

\bibitem{Davis10} T.-M. Chuang, M. P. Allan, J. Lee, Y. Xie, N. Ni,
S. L. Bud'ko, G. S. Boebinger, P. C. Canfield, and J. C. Davis, Science
\textbf{327}, 181 (2010); M. P. Allan \emph{et al}., unpublished.

\bibitem{Duzsa11} A. Dusza, A. Lucarelli, F. Pfuner, J.-H. Chu, I.
R. Fisher, and L. Degiorgi, EPL \textbf{93}, 37002 (2011); A. Dusza,
A. Lucarelli, A. Sanna, S. Massidda, J.-H. Chu, I.R. Fisher, and L.
Degiorgi, New J. Phys. \textbf{14} 023020 (2012). 

\bibitem{Uchida11} M. Nakajima, T. Liang, S. Ishida, Y. Tomioka,
K. Kihou, C. H. Lee, A. Iyo, H. Eisaki, T. Kakeshita, T. Ito, and
S. Uchida, PNAS \textbf{108}, 12238 (2011).

\bibitem{Shen11} M. Yi, D. Lu, J.-H Chu, J. G. Analytis, A. P. Sorini,
A. F. Kemper, B. Moritz, S.-K. Mo, R. G. Moore, M. Hashimoto, W.-S.
Lee, Z. Hussain, T. P. Devereaux, I. R. Fisher, and Z.-X. Shen, PNAS
\textbf{108}, 6878 (2011).

\bibitem{Song11} C.-L. Song, Y.-L. Wang, P. Cheng, Y.-P. Jiang, W.
Li, T. Zhang, Z. Li, K. He, L. Wang, J.-F. Jia, H.-H. Hung, C. Wu,
X. Ma, X. Chen, and Q.-K. Xue, Science \textbf{332}, 1410 (2011).

\bibitem{Matsuda11} Y. Matsuda \emph{et al}., unpublished.

\bibitem{Kuo11} H.-H. Kuo, J.-H. Chu, S. C. Riggs, L. Yu, P. L. McMahon,
K. De Greve, Y. Yamamoto, J. G. Analytis, and I. R. Fisher, Phys.
Rev. B \textbf{84}, 054540 (2011).

\bibitem{Blomberg12} E. C. Blomberg, M. A. Tanatar, R. M. Fernandes,
Bing Shen, Hai-Hu Wen, J. Schmalian, and R. Prozorov, arXiv:1202.4430.

\bibitem{kruger09} F. Krüger, S. Kumar, J. Zaanen, J. van den Brink,
Phys. Rev. B \textbf{79}, 054504 (2009).

\bibitem{RRPSingh09} R. R. P. Singh, arXiv:0903.4408.

\bibitem{w_ku10} C. C. Lee, W. G. Yin, and W. Ku, Phys. Rev. Lett.
\textbf{103}, 267001 (2009); W. G. Yin, C. C. Lee, and W. Ku, Phys.
Rev. Lett. \textbf{105}, 107004 (2010).

\bibitem{Phillips10} W. Lv, F. Krüger, and P. Phillips, Phys. Rev.
B \textbf{82}, 045125 (2010); W. Lv and P. Phillips, Phys. Rev. B
\textbf{84}, 174512 (2011).

\bibitem{devereaux10} C.-C. Chen, J. Maciejko, A. P. Sorini, B. Moritz,
R. R. P. Singh, and T. P. Devereaux, Phys. Rev. B \textbf{82}, 100504
(2010); R. Applegate, R. R. P. Singh, C.-C. Chen, and T. P. Devereaux,
Phys. Rev. B \textbf{85}, 054411 (2012).

\bibitem{Nevidomskyy11} A. H. Nevidomskyy, arXiv:1104.1747

\bibitem{bascones} E. Bascones, M. J. Calderon, and B. Valenzuela,
Phys. Rev. Lett. \textbf{104}, 227201 (2010); B. Valenzuela, E. Bascones,
and M. J. Calderon, Phys. Rev. Lett. \textbf{105}, 207202 (2010);
M. J. Calderon, G. Leon, B. Valenzuela, and E. Bascones, arXiv:1107.2279

\bibitem{dagotto} M. Daghofer, Q. Luo, R. Yu, D. Yao, A. Moreo, and
E. Dagotto, Phys. Rev. B \textbf{81}, 180514(R) (2010); P. M. R. Brydon,
M. Daghofer, and C. Timm, J. Phys.: Condens. Matter \textbf{23}, 246001
(2011); A. Nicholson, Q. Luo, W. Ge, J. Riera, M. Daghofer, G. B.
Martins, A. Moreo, and E. Dagotto, Phys. Rev. B \textbf{84}, 094519
(2011).

\bibitem{kotliar} Z. P. Yin, K. Haule, and G. Kotliar, Nature Phys.
\textbf{7}, 294 (2011).

\bibitem{Fang08} C. Fang, H. Yao, W.-F. Tsai, J. Hu, and S. A. Kivelson,
Phys. Rev. B \textbf{77,} 224509 (2008).

\bibitem{Xu08} C. Xu, M. Müller, and S. Sachdev, Phys. Rev. B \textbf{78},
020501(R) (2008).

\bibitem{FernandesPRL10} R. M. Fernandes, L. H. VanBebber, S. Bhattacharya,
P. Chandra, V. Keppens, D. Mandrus, M. A. McGuire, B. C. Sales, A.
S. Sefat, and J. Schmalian, Phys. Rev. Lett. \textbf{105}, 157003
(2010).

\bibitem{Si08} Q. Si and E. Abrahams, Phys. Rev. Lett. \textbf{101},
076401 (2008); E. Abrahams and Q. Si, J. Phys.: Condens. Matter \textbf{23},
223201 (2011); P. Goswami, R. Yu, Q. Si, and E. Abrahams, Phys. Rev.
B \textbf{84}, 155108 (2011).

\bibitem{Qi09} Y. Qi and C. Xu, Phys. Rev. B \textbf{80}, 094402
(2009).

\bibitem{Mazin09} I. I. Mazin and M. D. Johannes, Nature Phys. \textbf{5},
141 (2009).

\bibitem{Antropov11} A. L. Wysocki, K. D. Belashchenko, and V. P.
Antropov, Nat. Phys. \textbf{7}, 485 (2011).

\bibitem{JiangpingHu11} J. Hu, B. Xu, W. Liu, N. Hao, and Y. Wang,
arXiv:1106.5169

\bibitem{Batista11} Y. Kamiya, N. Kawashima, and C. D. Batista, Phys.
Rev. B \textbf{84}, 214429 (2011).

\bibitem{Lorenzana_nematic} M. Capati, M. Grilli, and J. Lorenzana,
Phys. Rev. B \textbf{84}, 214520 (2011).

\bibitem{FernandesPRB12} R. M. Fernandes, A. V. Chubukov, J. Knolle,
I. Eremin, and J. Schmalian, Phys. Rev. B \textbf{85}, 024534 (2012).

\bibitem{Levy71} P. M. Levy and H. H. Chen, Phys. Rev. Lett. \textbf{27},
1385 (1971).

\bibitem{Fradkin_review}  E. Fradkin, S. A. Kivelson, M. J. Lawler,
J. P. Eisenstein, and A. P. Mackenzie, Annu. Rev. Condens. Matter
Phys. \textbf{1}, 153 (2010).

\bibitem{IanLegendre}  J.-H. Chu, H.-H. Kuo, J. G. Analytis, and
I. R. Fisher, arXiv:1203.3239.

\bibitem{chandra} P. Chandra, P. Coleman, and A.I. Larkin, Phys.
Rev. Lett. \textbf{64}, 88 (1990).

\bibitem{Eremin10} I. Eremin and A. V. Chubukov, Phys. Rev. B \textbf{81},
024511 (2010).

\bibitem{comment} The saddle-point approximation takes into account,
in a self-consitent way, the fluctuations of the magnetic order parameter
$\mathbf{M}_{i}$, but not the fluctuations of the Ising order parameter
$\varphi$. It is formally valid in the limit $N\rightarrow\infty$,
where $N$ is the number of components of the magnetic order parameter
$\mathbf{M}_{i}$. For this reason, in the remainder of the text,
all relevant coupling constants $u$, $g$, $C_{s,0}$, etc are properly
renormalized such that the total action has an overall pre-factor
$N$.

\bibitem{Kim11} M. G. Kim, R. M. Fernandes, A. Kreyssig, J. W. Kim,
A. Thaler, S. L. Bud'ko, P. C. Canfield, R. J. McQueeney, J. Schmalian,
and A. I. Goldman, Phys. Rev. B \textbf{83}, 134522 (2011).

\bibitem{Birgeneau11} C. R. Rotundu and R. J. Birgeneau, Phys. Rev.
B \textbf{84}, 092501 (2011).

\bibitem{Hu12} J. Hu, C. Setty and S. Kivelson, arXiv:1201.5174.

\bibitem{Cano12} A. Cano and I. Paul, arXiv:1201.5594.

\bibitem{Blomberg11} E. C. Blomberg, A. Kreyssig, M. A. Tanatar,
R. M. Fernandes, M. G. Kim, A. Thaler, J. Schmalian, S. L. Bud'ko,
P. C. Canfield, A. I. Goldman, and R. Prozorov, arXiv:1111.0997.

\bibitem{Dhital12} C. Dhital, Z. Yamani, W. Tian, J. Zeretsky, A.
S. Sefat, Z. Wang, R. J. Birgeneau, and S. D. Wilson, Phys. Rev. Lett.
\textbf{108}, 087001 (2012).

\bibitem{Ma_NaFeAs} L. Ma, G. F. Chen, D.-X. Yao, J. Zhang, S. Zhang,
T.-L. Xia, and W. Yu, Phys. Rev. B \textbf{83}, 132501 (2011).

\bibitem{kitagawa_NaFeAs} K. Kitagawa, Y. Mezaki, K. Matsubayashi,
Y. Uwatoko, and M. Takigawa, J. Phys. Soc. Jpn. \textbf{80}, 033705
(2011).

\bibitem{Diallo10} S. O. Diallo, D. K. Pratt, R.M. Fernandes, W.
Tian, J. L. Zarestky, M. Lumsden, T. G. Perring, C. L. Broholm, N.
Ni, S. L. Bud'ko, P. C. Canfield, H.-F. Li, D. Vaknin, A. Kreyssig,
A .I. Goldman, and R. J. McQueeney, Phys. Rev. B \textbf{81}, 214407
(2010).

\bibitem{Li10} H.-F. Li, C. Broholm, D. Vaknin, R. M. Fernandes,
D. L. Abernathy, M. B. Stone, D. K. Pratt, W. Tian, Y. Qiu, N. Ni,
S. O. Diallo, J. L. Zarestky, S. L. Bud'ko, P. C. Canfield, and R.
J. McQueeney, Phys. Rev. B \textbf{82}, 140503(R) (2010).

\bibitem{magn_pseudogap} M. Vilk and A.-M. S. Tremblay, J. Phys.
I France \textbf{7}, 1309-1368 (1997); J. Schmalian, D. Pines, and
B. Stojkovi\'{c}, Phys. Rev. Lett. \textbf{80}, 3839 (1998); Phys.
Rev. B \textbf{60}, 667 (1999); E.Z. Kuchinskii and M. V. Sadovskii,
JETP \textbf{88}, 968 (1999); T. Sedrakyan and A.V. Chubukov, Phys.
Rev B \textbf{81}, 174536 (2010).

\bibitem{Tanatar_pseudogap} M. A. Tanatar, N. Ni, A. Thaler, S. L.
Bud'ko, P. C. Canfield, and R. Prozorov, Phys. Rev. B \textbf{82},
134528 (2010).

\bibitem{Xu_pseudogap} Y.-M. Xu, P. Richard, K. Nakayama, T. Kawahara,
Y. Sekiba, T. Qian, M. Neupane, S. Souma, T. Sato, T. Takahashi, H.-Q.
Luo, H.-H. Wen, G.-F. Chen, N.-L. Wang, Z. Wang, Z. Fang, X. Dai,
and H. Ding, Nature Comm. \textbf{2}, 392 (2011).

\bibitem{Fernandes11} R. M. Fernandes, E. Abrahams, and J. Schmalian,
Phys. Rev. Lett. \textbf{107}, 217002 (2011).

\bibitem{Hlubina95} R. Hlubina and T. M. Rice, Phys. Rev. B \textbf{51},
9253 (1995).

\bibitem{Rosch99} A. Rosch, Phys. Rev. Lett. \textbf{82}, 4280 (1999).

\bibitem{Liu10} C. Liu, T. Kondo, R. M. Fernandes, A. D. Palczewski,
E. D. Mun, N. Ni, A. N. Thaler, A. Bostwick, E. Rotenberg, J. Schmalian,
S. L. Bud'ko, P. C. Canfield, and A. Kaminski, Nature Phys. \textbf{6},
419 (2010).

\bibitem{Dagotto_Liang} S. Liang, G. Alvarez, C. Sen, A. Moreo, and
E. Dagotto, arXiv:1111.6994

\bibitem{Yoshizawa12} M. Yoshizawa, D. Kimura, T. Chiba, A. Ismayil,
Y. Nakanishi, K. Kihou, C.-H. Lee, A. Iyo, H. Eisaki, M. Nakajima,
and S. Uchida, J. Phys. Soc. Jpn. \textbf{81}, 024604 (2012).

\bibitem{Wang08} W. Z. Hu, J. Dong, G. Li, Z. Li, P. Zheng, G. F.
Chen, J. L. Luo, and N. L. Wang, Phys. Rev. Lett. \textbf{101}, 257005
(2008).

\bibitem{ZXShen_NaFeAS} M. Yi, D. H. Lu, R. G. Moore, K. Kihou, C.-H.
Lee, A. Iyo, H. Eisaki, T. Yoshida, A. Fujimori, and Z.-X. Shen, arXiv:1111.6134.

\bibitem{Zhang09} J. Zhang, R. Sknepnek, R. M. Fernandes, and J.
Schmalian, Phys. Rev. B \textbf{79}, 220502(R) (2009).

\bibitem{Fernandes_Schmalian} R. M. Fernandes and J. Schmalian, Phys.
Rev. B \textbf{82}, 014520 (2010); \emph{ibid} Phys. Rev. B \textbf{82},
014521 (2010).

\bibitem{Vorontsov10} A. B. Vorontsov, M. G. Vavilov, and A. V. Chubukov,
Phys. Rev. B \textbf{79}, 060508(R) (2009); \emph{ibid} Phys. Rev.
B \textbf{81}, 174538 (2010).

\bibitem{Moon11} E. G. Moon and S. Sachdev, arXiv:1112.3973.

\bibitem{Abanov99} Ar. Abanov and A. V. Chubukov, Phys. Rev. Lett.
\textbf{83}, 1652 (1999).

\bibitem{maiti} S. Maiti, M. M. Korshunov, T. A. Maier, P. J. Hirschfeld,
and A. V. Chubukov, Phys. Rev. B \textbf{84}, 224505 (2011); Phys.
Rev. Lett. \textbf{107}, 147002 (2011).

\bibitem{orbital} S. Onari and H. Kontani, Phys. Rev. Lett. \textbf{103},
177001 (2009); H. Kontani and S. Onari, Phys. Rev. Lett. \textbf{104},
157001 (2010); Y. Yanagi, Y. Yamakawa, and Y. Ono, Phys. Rev. B \textbf{81},
054518 (2010); T. Saito, S. Onari, and H. Kontani Phys. Rev. B \textbf{83},
140512(R) (2011).

\bibitem{Christianson08} A. D. Christianson, E. A. Goremychkin, R.
Osborn, S. Rosenkranz, M. D. Lumsden, C. D. Malliakas, I. S. Todorov,
H. Claus, D. Y. Chung, M. G. Kanatzidis, R. I. Bewley, and T. Guidi,
Nature \textbf{456}, 930 (2008).

\bibitem{Hanaguri10} T. Hanaguri, S. Niitaka, K. Kuroki, and H. Takagi,
Science \textbf{23}, 474 (2010).

\bibitem{flux_jumps} C.-T. Chen, C. C. Tsuei, M. B. Ketchen, Z.-A.
Ren, and Z. X. Zhao, Nature Phys. \textbf{6}, 260 (2010).

\bibitem{LiFeAs_gap} S. V. Borisenko, V. B. Zabolotnyy, A. A. Kordyuk,
D. V. Evtushinsky, T. K. Kim, I. V. Morozov, R. Follath, and B. Büchner,
Symmetry \textbf{4}, 251 (2012); M. P. Allan, A. W. Rost, A. P. Mackenzie,
Yang Xie, J. C. Davis, K. Kihou, C. H. Lee, A. Iyo, H. Eisaki, and
T.-M. Chuang, Science \textbf{336}, 563 (2012). 

\bibitem{Abanov01}  Ar. Abanov, A. V. Chubukov, and J. Schmalian,
Europhys. Lett. \textbf{55}, 369 (2001).

\bibitem{zlatko} J. Kang and Z. Tesanovic, Phys. Rev. B \textbf{83},
020505 (2011).

\bibitem{Vojta10} A. Hackl and M. Vojta, New J. Phys. \textbf{11},
055064 (2009).

\bibitem{Laad11} M. S. Laad and L. Craco, Phys. Rev. B \textbf{84},
054530 (2011).

\bibitem{kontani_nematic} H. Kontani, T. Saito, and S. Onari, Phys.
Rev. B \textbf{84}, 024528 (2011); Y. Inoue, Y. Yamakawa, and H. Kontani,
arXiv:1110.2401.

\bibitem{daghofer_nematic} M. Daghofer, A. Nicholson, and A. Moreo,
arXiv:1202.3656.

\bibitem{Indranil11} I. Paul, Phys. Rev. Lett. \textbf{107}, 047004
(2011).

\bibitem{Cano10} A. Cano, M. Civelli, I. Eremin, and I. Paul, Phys.
Rev. B \textbf{82}, 020408(R) (2010).

\bibitem{Sachdev_nematic11} D. Chowdhury, E. Berg, and S. Sachdev,
Phys. Rev. B \textbf{84}, 205113 (2011).

\bibitem{cano_domains} A. Cano, Phys. Rev. B \textbf{84}, 012504
(2011).

\bibitem{kivelson_cuprates} S. A. Kivelson, E. Fradkin, V. Oganesyan,
I. P. Bindloss, J. M. Tranquada, A. Kapitulnik, and C. Howald, Rev.
Mod. Phys. \textbf{75}, 1201 (2003).

\bibitem{matsuda_hidden_order} R. Okazaki, T. Shibauchi, H. J. Shi,
Y. Haga, T. D. Matsuda, E. Yamamoto, Y. Onuki, H. Ikeda, and Y. Matsuda,
Science \textbf{331}, 439 (2011). 
\end{thebibliography}
\end{document}